\pdfoutput=1

\documentclass[11pt]{article}
\usepackage{emnlp2021}
\usepackage{times}
\usepackage{latexsym}
\usepackage{url}
\usepackage{graphics}
\usepackage{amsmath}
\usepackage{amssymb}
\usepackage{amsfonts}
\usepackage{mathtools}
\usepackage{booktabs}
\usepackage{subcaption}
\usepackage{siunitx}
\usepackage{booktabs}
\usepackage{colortbl}
\usepackage{array}
\usepackage{soul}
\usepackage{pifont}
\usepackage{tabularx}
\usepackage{multirow}
\usepackage{tikz-dependency}
\usepackage[linesnumbered,ruled,vlined]{algorithm2e}
\usepackage{textcomp}
\usepackage{pifont}

\usepackage[T1]{fontenc}
\usepackage[utf8]{inputenc}

\usepackage{microtype}

\newcommand{\rmn}{APRE}  
\newcommand{\tmn}{ASPE}  

\newcommand{\numbl}{Thirteen}

\newcommand{\numdsl}{seven}

\usepackage{xcolor}
\newcommand{\td}[1]{{\bf\color{red}[{\sc TODO:} #1]}}

\newcommand{\SM}{\textbf{Supplementary Materials}}

\newcommand{\matA}{\boldsymbol{\mathrm{A}}}

\newcommand{\matW}{\boldsymbol{\mathrm{W}}}

\newcommand{\matH}{\boldsymbol{\mathrm{H}}}

\newcommand{\matG}{\boldsymbol{\mathrm{G}}}

\newcommand{\veca}{\boldsymbol{a}}
\newcommand{\vecg}{\boldsymbol{g}}

\newcommand{\vecw}{\boldsymbol{w}}

\newcommand{\vecb}{\boldsymbol{b}}

\newcommand{\vech}{\boldsymbol{h}}
\newcommand{\vecv}{\boldsymbol{v}}

\newcommand{\vecgamma}{\boldsymbol{\gamma}}

\newcommand{\myTitle}{Recommend for a Reason: Unlocking the Power of Unsupervised Aspect-Sentiment Co-Extraction}

\title{\myTitle}

\author{%
Zeyu Li$^1$, Wei Cheng$^2$, Reema Kshetramade$^1$, John Houser$^1$, Haifeng Chen$^2$, and Wei Wang$^1$\\
$^1$Department of Computer Science, University of California, Los Angeles\\
$^2$NEC Labs America\\
 \texttt{\{zyli,weiwang\}@cs.ucla.edu}\\ 
 \texttt{\{reemakshe,johnhouser\}@ucla.edu} \\
 \texttt{\{weicheng,haifeng\}@nec-labs.com}
}

\date{}

\begin{document}
\maketitle

\begin{abstract}
Compliments and concerns in reviews are valuable for understanding users' shopping interests and their opinions with respect to specific aspects of certain items. 
Existing review-based recommenders favor large and complex language encoders that can only learn latent and uninterpretable text representations.
They lack explicit user-attention and item-property modeling, which however could provide valuable information beyond the ability to recommend items. Therefore, we propose a tightly coupled two-stage approach, including an Aspect-Sentiment Pair Extractor (\tmn{}) and an Attention-Property-aware Rating Estimator (\rmn{}). Unsupervised \tmn{} mines Aspect-Sentiment pairs (AS-pairs) and \rmn{} predicts ratings using AS-pairs as concrete aspect-level evidences. Extensive experiments on \numdsl{} real-world Amazon Review Datasets demonstrate that ASPE can effectively extract AS-pairs which enable \rmn{} to deliver superior accuracy over the leading baselines.

\end{abstract}

\definecolor{rulecolor}{RGB}{0,0,0}
\definecolor{tableheadcolor}{gray}{0.92}
%
\newcommand{\topline}{ %
        \arrayrulecolor{rulecolor}\specialrule{0.1em}{\abovetopsep}{0pt}%
        \arrayrulecolor{tableheadcolor}\specialrule{\belowrulesep}{0pt}{0pt}%
        \arrayrulecolor{rulecolor}}

\newcommand{\midtopline}{ %
        \arrayrulecolor{tableheadcolor}\specialrule{\aboverulesep}{0pt}{0pt}%
        \arrayrulecolor{rulecolor}\specialrule{\lightrulewidth}{0pt}{0pt}%
        \arrayrulecolor{white}\specialrule{\belowrulesep}{0pt}{0pt}%
        \arrayrulecolor{rulecolor}}

\newcommand{\bdm}[0]{\textbf{\dag}} 

\section{Introduction}
\label{sec:intro}



Reviews and ratings are valuable assets for the recommender systems of e-commerce websites since they immediately describe the users' subjective feelings about the purchases. Learning user preferences from such feedback is straightforward and efficacious.
Previous research on review-based recommendation has been fruitful~\cite{chin2018anr,chen2018neural,bauman2017aspect,liu2019daml}. 
Cutting-edge natural language processing (NLP) techniques are applied to extract the latent user sentiments, item properties, and the complicated interactions between the two components.

However, existing approaches have disadvantages bearing room for improvement. Firstly, they dismiss the phenomenon that users may hold different \textit{attentions} toward various \textit{properties} of the merchandise. 
An item property is the combination of an aspect of the item and the characteristic associated with it.
Users may show strong attentions to certain properties but indifference to others. 
The attended advantageous or disadvantageous properties can dominate the \textit{attitude} of users and consequently, decide their generosity in rating.
\newcommand{\mygreen}[1]{\textcolor[HTML]{2E9E2E}{#1}}  
\newcommand{\myred}[1]{\textcolor[HTML]{CC0000}{#1}}  
\newcommand{\myblue}[1]{\textcolor[HTML]{0066FF}{#1}}  

\newcommand{\reviewA}{\textbf{R1 [5 stars]: }\textit{\mygreen{Comfortable.} \myred{Very high quality sound.} \dots \myblue{Mic is good too.} There is an switch to mute your mic\dots I wear glasses and \mygreen{these are comfortable with my glasses on.} \dots}}
\newcommand{\reviewB}{\textbf{R2 [3 stars]: }\textit{\mygreen{I love the comfort,} \myred{sound}, and style but \myblue{the mic is complete junk!}}}
\newcommand{\reviewC}{\textbf{R3 [5 stars]: }\textit{\dots But this one \mygreen{feels like a pillow,} \myred{there's nothing wrong with the audio} and it does the job. \dots con is that \myblue{the included microphone is pretty bad.}}}

\newcolumntype{A}{>{\centering\arraybackslash}p{0.1\linewidth}}
\newcolumntype{R}{>{\raggedright\arraybackslash}p{0.6\linewidth}}

\begin{table*}[!ht]
\centering
\small
 \setlength\tabcolsep{3pt}
\resizebox{0.99\linewidth}{!}{
\begin{tabularx}{\linewidth}{X|AAA}
\topline
\rowcolor[gray]{0.92}
\centering{\textbf{Reviews}} & \textbf{\myblue{Microphone}} & \textbf{\mygreen{Comfort}} &\myred{\textbf{Sound}} \\
\midtopline
\reviewA & \textit{good} (satisfied) & \textit{comfortable} & \textit{high quality} (praising) \\
\midrule
\reviewB & \textit{complete junk} (angry) & \textit{love} & \textit{love} \\
\midrule
\reviewC & \textit{pretty bad} (unsatisfied) & \textit{like a pillow} (enjoyable) & \textit{nothing wrong } \\
\bottomrule
\end{tabularx}
}
\caption{Example reviews of a headset with three aspects, namely \myblue{microphone quality}, \mygreen{comfort level}, and \myred{sound quality}, highlighted specifically. The extracted sentiments are on the right. R1 vs. R2: Different users react differently (microphone quality) to the same item due to distinct personal attentions and, consequently, give divergent ratings. R1 vs. R3: A user can still rate highly of an item due to special attention on particular aspects (comfort level) regardless of certain unsatisfactory or indifferent properties (microphone and sound qualities).}
\label{tab:review_examples}
\end{table*}

Table~\ref{tab:review_examples} exemplifies the impact of the user attitude using three real reviews for a headset. Three aspects are covered: microphone quality, comfortableness, and sound quality. 
The microphone quality is controversial. R2 and R3 criticize it but R1 praises it. 
The sole disagreement between R1 and R2 is on microphone, which is the major concern of R2, results in the divergence of ratings (5 stars vs. 3 stars). 
However, 
R3 neglects that disadvantage and grades highly (5 stars) for its superior comfortableness indicated by the metaphor of ``pillow''.

Secondly, understanding user motivations in granular item properties provides valuable information beyond the ability to recommend items. It requires aspect-based NLP techniques to extract explicit and definitive aspects.
However, existing aspect-based models mainly use \textit{latent} or \textit{implicit} aspects~\cite{chin2018anr} whose real semantics are unjustifiable.
Similar to Latent Dirichlet Allocation (LDA, ~\citealp{blei2003latent}), the semantics of the derived aspects (topics) are mutually overlapped~\cite{huang2020weakly}.
These models undermine the resultant aspect distinctiveness and lead to uninterpretable and sometimes counterintuitive results.
The root of the problem is the lack of large review corpora with aspect and sentiment annotations. The existing ones are either too small or too domain-specific~\cite{wang2018recursive} to be applied to general use cases. 
Progress on sentiment term extraction~\cite{dai2019rinante,tian2020skep,chen2020synchronous} 
takes advantage of neural networks and linguistic knowledge and
partially makes it possible to use unsupervised term annotation to tackle the lack-of-huge-corpus issue.


In this paper, we seek to understand how reviews and ratings are affected by users' perception of item properties in a fine-grained way and discuss how to utilize these findings transparently and effectively in \textit{rating prediction}. 
We propose a two-stage recommender with an \textbf{unsupervised} Aspect-Sentiment Pair Extractor (\tmn{}) and an Attention-Property-aware Rating Estimator (\rmn{}). 
\tmn{} extracts \texttt{(aspect, sentiment)} pairs (AS-pairs) from reviews.
The pairs are fed into \rmn{} as \textit{explicit} user attention and item property carriers indicating both \textit{frequencies} and \textit{sentiments} of aspect mentions.
\rmn{} encodes the text by a contextualized encoder and processes \textit{implicit} text features and the annotated AS-pairs by a dual-channel rating regressor.
\tmn{} and \rmn{} jointly extract explicit aspect-based attentions and properties and solve the rating prediction with a great performance.

Aspect-level user attitude differs from \textit{user preference}.
The user attitudes produced by the interactions of user attentions and item properties are sophisticated and granular sentiments and rationales for interpretation (see Section~\ref{sec:case_study} and~\ref{subsec:case_study_ii}).
Preferences, on the contrary, are coarse sentiments such as like, dislike, or neutral. 
Preference-based models may infer that R1 and R3 are written by headset lovers because of the high ratings. 
Instead, attitude-based methods further understand that it is the comfortableness that matters to R3 rather than the item being a headset. 
Aspect-level attitude modeling is more accurate, informative, and personalized than preference modeling. 

\textbf{Note.} Due to the page limits, some supportive materials, marked by ``\bdm{}'', are presented in the \SM{}. We strongly recommend readers check out these materials.
The source code of our work is available on GitHub at~\url{https://github.com/zyli93/ASPE-APRE}.
\section{Related Work}
\label{sec:relatedwork}
Our work is related to four lines of literature which are located in the overlap of ABSA and Recommender Systems.

\subsection{Aspect-based Sentiment Analysis}
Aspect-based sentiment analysis (ABSA)~\cite{dombert,wang2018target} predicts sentiments toward aspects mentioned in the text.
Natural language is modeled by graphs in \cite{gcnsa,rgatSA} such as Pointwise Mutual Information (PMI) graphs and dependency graphs.
\citet{phan2020modelling} and~\citet{tang2020dependency} utilize contextualized language encoding to capture the context of aspect terms.
\citet{chen2020aspect} focuses on the consistency of the emotion surrounding the aspects, and~\citet{du2020adversarial} equips pre-trained BERT with domain-awareness of sentiments. Our work is informed by these progress which utilize PMI, dependency tree, and BERT for syntax feature extraction and language encoding.

\subsection{Aspect or Sentiment Terms Extraction}
Aspect and sentiment terms extraction is a presupposition of ABSA. However, manually annotating data for training, which requires the hard labor of experts, is only feasible on small datasets in particular domains such as Laptop and Restaurant~\cite{semeval14,semeval15} which are overused in ABSA.

Recently, RINANTE~\cite{dai2019rinante} and SDRN~\cite{chen2020synchronous} automatically extract both terms using rule-guided data augmentation and double-channel opinion-relation co-extraction, respectively.
However, the supervised approaches are too domain-specific to generalize to out-of-domain or open-domain corpora. Conducting domain adaptation from small labeled corpora to unlabeled open corpora only produces suboptimal results~\cite{wang2018recursive}.
SKEP~\cite{tian2020skep} exploits an unsupervised PMI+seed strategy to coarsely label sentimentally polarized tokens as sentiment terms, showing that the unsupervised method is advantageous when annotated corpora are insufficient in the domain-of-interest.

Compared to the above models, our \tmn{} has two merits of being (1) unsupervised and hence free from expensive data labeling; (2) generalizable to different domains by combining three different labeling methods. 


\subsection{Aspect-based Recommendation}
Aspect-based recommendation is a relevant task with a major difference that specific terms indicating sentiments are not extracted. Only the aspects are needed~\cite{hou2019explainable,guan2019attentive,huang2020personalized,chin2018anr}. Some disadvantages are summarized as follows. Firstly, the aspect extraction tools are usually outdated and inaccurate such as LDA~\cite{hou2019explainable}, TF-IDF~\cite{guan2019attentive}, and word embedding-based similarity~\cite{huang2020personalized}. Second, the representation of sentiment is scalar-based which is coarser than embedding-based used in our work. 

\subsection{Rating Prediction}
Rating prediction is an important task in recommendation. Related approaches utilize text mining algorithms to build user and item representations and predict ratings~\cite{kim2016convolutional,zheng2017joint,chen2018neural,chin2018anr,liu2019daml,bauman2017aspect}. 
However, the text features learned are latent and unable to provide explicit hints for explaining user interests. 
\section{\tmn{} and \rmn{}}
\label{sec:method}

\subsection{Problem Formulation}
\label{sec:problem_formulation}
Review-based rating prediction involves two major entities: users and items.
A user $u$ writes a review $r_{u,t}$ for an item $t$ and rates a score $s_{u,t}$. Let $R^u$ denote all reviews given by $u$ and $R^t$ denote all reviews received by $t$.
A rating regressor takes in a tuple of a review-and-rate event $(u, t)$ and review sets $R^u$ and $R^t$ to estimate the rating score $s_{u,t}$.

\subsection{Unsupervised \tmn{}}
\begin{figure}[!h]
    \centering
    \includegraphics[width=0.8\linewidth]{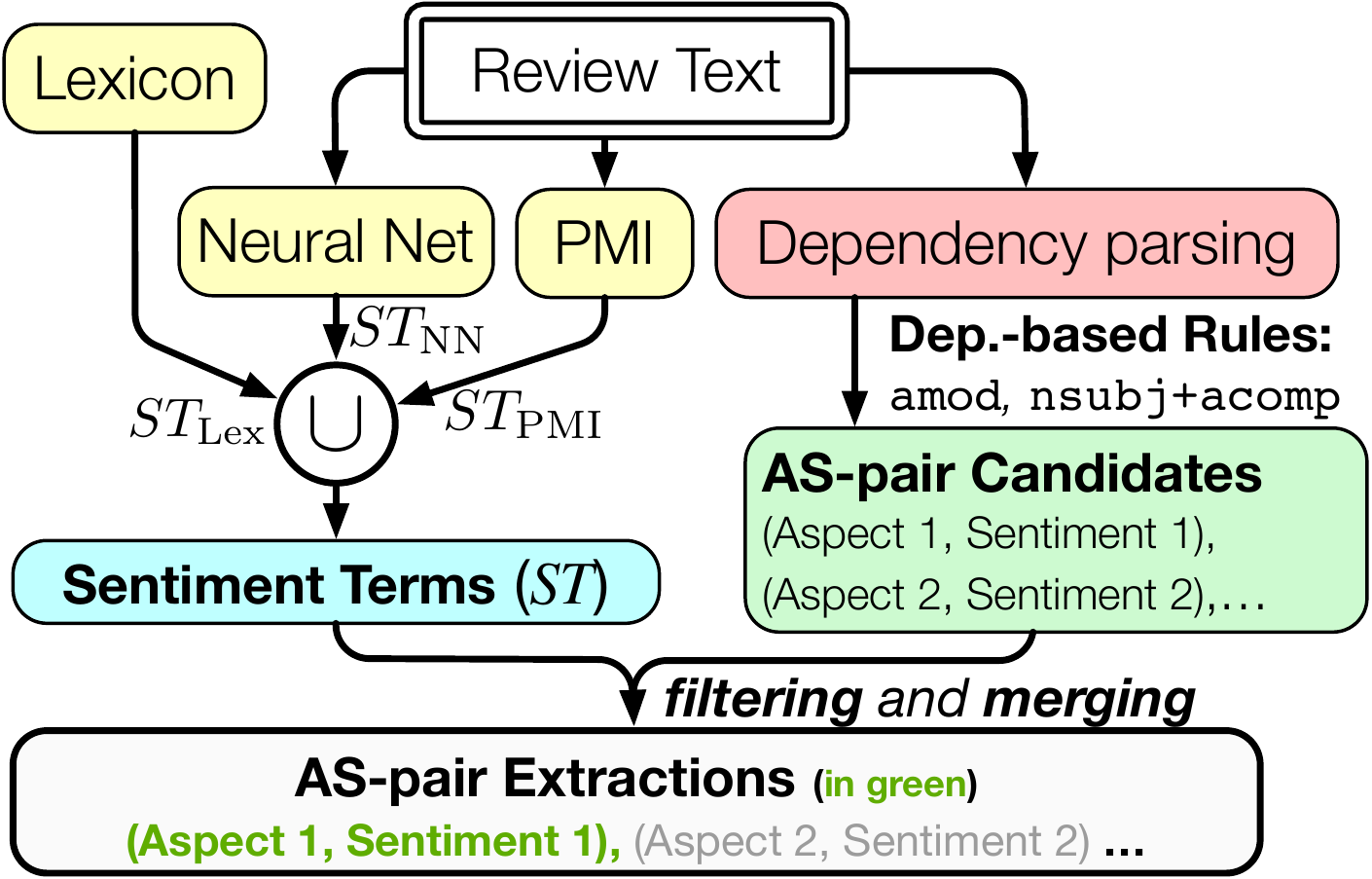}
    \caption{Pipeline of \tmn{}.}
    \label{fig:part_one_pipeline}
\end{figure}

We combine three separate methods to label AS-pairs without the need for supervision, namely PMI-based, neural network-based (NN-based), and language knowledge- or lexicon-based methods. The framework is visualized in Figure~\ref{fig:part_one_pipeline}.
\label{subsec:extractor}

\subsubsection{Sentiment Terms Extraction}
\label{subsec:senti_term_extraction}
\paragraph{PMI-based method} Pointwise Mutual Information (PMI) originates from Information Theory and is adapted into NLP~\cite{gcnsa,tian2020skep} to measure statistical word associations in corpora. It determines the sentiment polarities of words using a small number of carefully selected positive and negative seeds ($s^+$ and $s^-$)~\cite{tian2020skep}. It first extracts candidate sentiment terms satisfying the part-of-speech patterns by~\citet{turney2002} and then measures the polarity of each candidate term $w$ by
\begin{equation}
    \text{Pol}(w)=\sum_{s^+}\text{\small{PMI}}(w,s^+) - \sum_{s^-}\text{\small{PMI}}(w,s^-).
    \label{eq:polarity}
\end{equation}
Given a sliding window-based context sampler $ctx$,
the PMI$(\cdot,\cdot)$ between words is defined by 
\begin{equation}
    \text{PMI}(w_1,w_2)=\log\frac{p(w_1, w_2)}{p(w_1)p(w_2)},
    \label{eq:pmi_mat}
\end{equation}
where $p(\cdot)$, the probability estimated by token counts, is defined by $p(w_1,w_2)=\frac{|\{ctx|w_1, w_2 \in ctx\}|}{\text{total }\#ctx}$ and $p(w_1)=\frac{|\{ctx|w_1\in ctx\}|}{\text{total }\#ctx}$.
Afterward, we collect the top-$q$ sentiment tokens with strong polarities, both positive and negative, as $ST_\text{PMI}$.

\paragraph{NN-based method} As discussed in Section~\ref{sec:relatedwork}, co-extraction models~\cite{dai2019rinante} can accurately label AS-pairs only in the training domain. For sentiment terms with consistent semantics in different domains such as \textit{good} and \textit{great}, NN methods can still provide a robust extraction recall. In this work, we take a pretrained SDRN~\cite{chen2020synchronous} as the NN-based method to generate $ST_\text{NN}$. The pretrained SDRN is considered an off-the-shelf tool similar to the pretrained BERT which is \textbf{irrelevant} to our rating prediction data. Therefore, we argue \tmn{} is unsupervised for open domain rating prediction.
\paragraph{Knowledge-based method} PMI- and NN-based methods have shortcomings. The PMI-based method depends on the seed selection. The accuracy of the NN-based method deteriorates when the applied domain is distant from the training data. As compensation, we integrate a sentiment lexicon $ST_\text{Lex}$ summarized by linguists since expert knowledge is widely used in unsupervised learning. 
Examples of linguistic lexicons include SentiWordNet~\cite{baccianella2010sentiwordnet} and Opinion Lexicon~\cite{hu2004mining}. The latter one is used in this work.

\paragraph{Building sentiment term set} The three sentiment term subsets are joined to build an overall sentiment set used in AS-pair generation: $ST = ST_\text{PMI} \cup ST_\text{NN} \cup ST_\text{Lex}.$ The three sets compensate for the discrepancies of other methods and expand the coverage of terms shown in Table~\ref{tab:three_senti_terms}\bdm.

\subsubsection{Syntactic AS-pairs Extraction}
\label{subsec:aspair_generation}
To extract AS-pairs, we first label AS-pair \textit{candidates} using dependency parsing and then filter out non-sentiment-carrying candidates using ($ST$)\footnote{Section~\ref{subsubsec:pseudocode}\bdm{} explains this procedure in detail by pseudocode of Algorithm~\ref{alg:pseudocode_aspair_building}\bdm.}.
Dependency parsing extracts the syntactic relations between the words. Some nouns are considered potential aspects and are modified by adjectives with two types of dependency relations shown in Figure~\ref{tab:dep_rules}: \texttt{amod} and \texttt{nsubj+acomp}. The pairs of nouns and the modifying adjectives compose the AS-pair candidates. Similar techniques are widely used in unsupervised aspect extraction models~\cite{cat,dai2019rinante}.
AS-pair candidates are noisy since not all adjectives in it bear sentiment inclination. $ST$ comes into use to \textbf{filter} out non-sentiment-carrying AS-pair candidates whose adjective is not in $ST$. The left candidates form the AS-pair set. 
Admittedly, the dependency-based extraction for (noun, adj.) pairs is \textit{suboptimal} and causes missing aspect or sentiment terms. An implicit module is designed to remedy this issue. 
Open domain AS-pair co-extraction is blocked by the lacking of public labeled data and is left for future work.

We introduce \texttt{ItemTok} as a special aspect token of the \texttt{nsubj+acomp} rule where \texttt{nsubj} is a pronoun of the item such as \textit{it} and \textit{they}.
Infrequent aspect terms with less than $c$ occurrences are ignored to reduce sparsity.
We use WordNet synsets~\cite{miller1995wordnet} to \textbf{merge} the synonym aspects. The aspect with the most synonyms is selected as the representative of that aspect set.
\newcommand{\egamod}{
    \resizebox{1.2\linewidth}{!}{%
        \begin{dependency}[edge style={black,very thick},
        label style={scale=1.3}]
            \begin{deptext}[column sep=.5cm, row sep=.1ex]
            Amazing \& sound \& and \& quality, \& all \& in \& one \& headset. \\
            \end{deptext}
            \depedge[label style={fill=orange!60}]{2}{1}{amod}
            \depedge{2}{3}{cc}
            \depedge[label style={fill=purple!30}]{2}{4}{conj}
            \depedge{2}{6}{prep}
            \depedge{6}{5}{advmod}
            \depedge{6}{8}{pobj}
            \depedge{8}{7}{nummod}
            \wordgroup[group style={fill=yellow!40, draw=yellow!40}]{1}{2}{2}{aspect1}
            \wordgroup[group style={fill=yellow!40, draw=yellow!40}]{1}{4}{4}{aspect2}
            \wordgroup[group style={fill=green!50, draw=green!50}]{1}{1}{1}{sentiment1}
        \end{dependency}
    }
}

\newcommand{\egacomp}{
    \resizebox{1.2\linewidth}{!}{%
        \begin{dependency}[edge style={black,very thick},
        label style={scale=1.3}]
            \begin{deptext}[column sep=.5cm, row sep=.1ex]
            Sound \& quality \& is \& superior \& and \& comfort \& is \& excellent. \\
            \end{deptext}
            \depedge[label style={fill=purple!30}]{2}{1}{compound}
            \depedge[label style={fill=cyan!30}]{3}{2}{nsubj}
            \depedge[label style={fill=red!50}]{3}{4}{acomp}
            \depedge{3}{5}{cc}
            \depedge{3}{7}{conj}
            \depedge[label style={fill=cyan!30}]{7}{6}{nsubj}
            \depedge[label style={fill=red!50}]{7}{8}{acomp}
            \wordgroup[group style={fill=yellow!40, draw=yellow!40}]{1}{1}{2}{aspect1}
            \wordgroup[group style={fill=yellow!40, draw=yellow!40}]{1}{6}{6}{aspect2}
            \wordgroup[group style={fill=green!50, draw=green!50}]{1}{4}{4}{sentiment1}
            \wordgroup[group style={fill=green!50, draw=green!50}]{1}{8}{8}{sentiment2}
        \end{dependency}
    }
}

\begin{table}[!h]
    \centering
    \small
    \resizebox{.95\linewidth}{!}{%
    \begin{tabular}{l}
        \toprule
        \textbf{\texttt{amod} dependency relation:} \\
        \egamod \\
        \textbf{Extracted AS-pair candidates:} \\
        \quad \texttt{(sound, amazing)}, \texttt{(quality, amazing)} \\
        \midrule
        \textbf{\texttt{nsubj}+\texttt{acomp} dependency relation:} \\
        \egacomp \\
        \textbf{Extracted AS-pair candidates:} \\
        \quad \texttt{(Sound quality, superior)}, \texttt{(comfort, excellent)} \\
        \bottomrule
    \end{tabular}
    }
    
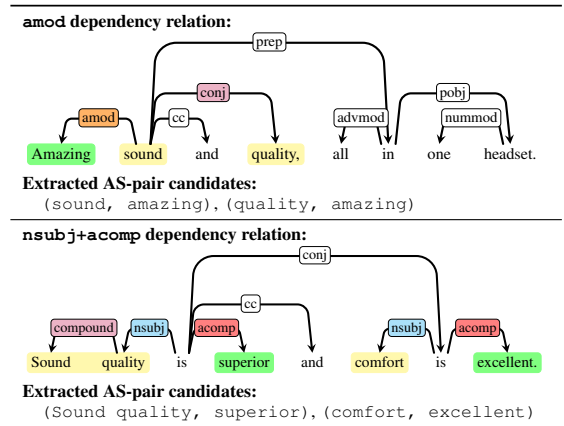
\captionof{figure}{Two dependency-based rules for AS-pair candidates extraction. Effective dependency relations and \colorbox{yellow!40}{aspects} and \colorbox{green!50}{sentiments} \textit{candidates} are highlighted.}
    \label{tab:dep_rules}
    \vspace{-1.5em}
\end{table}

\paragraph{Discussion} \tmn{} is different from \textit{Aspect Extraction} (AE)~\cite{cat,doer,wei2020don,ma2019exploring,angelidis2018summariing,doubleembedding,lifelong,he2017unsupervised} which extracts aspects only and infers sentiment polarities in \{\texttt{pos}, \texttt{neg}, (\texttt{neu})\}. AS-pair co-extraction, however, offers more diversified emotional signals than the bipolar sentiment measurement of AE.

\subsection{\rmn{}}
\label{subsec:rmn} 
\begin{figure}
    \centering
    \includegraphics[width=0.9\linewidth]{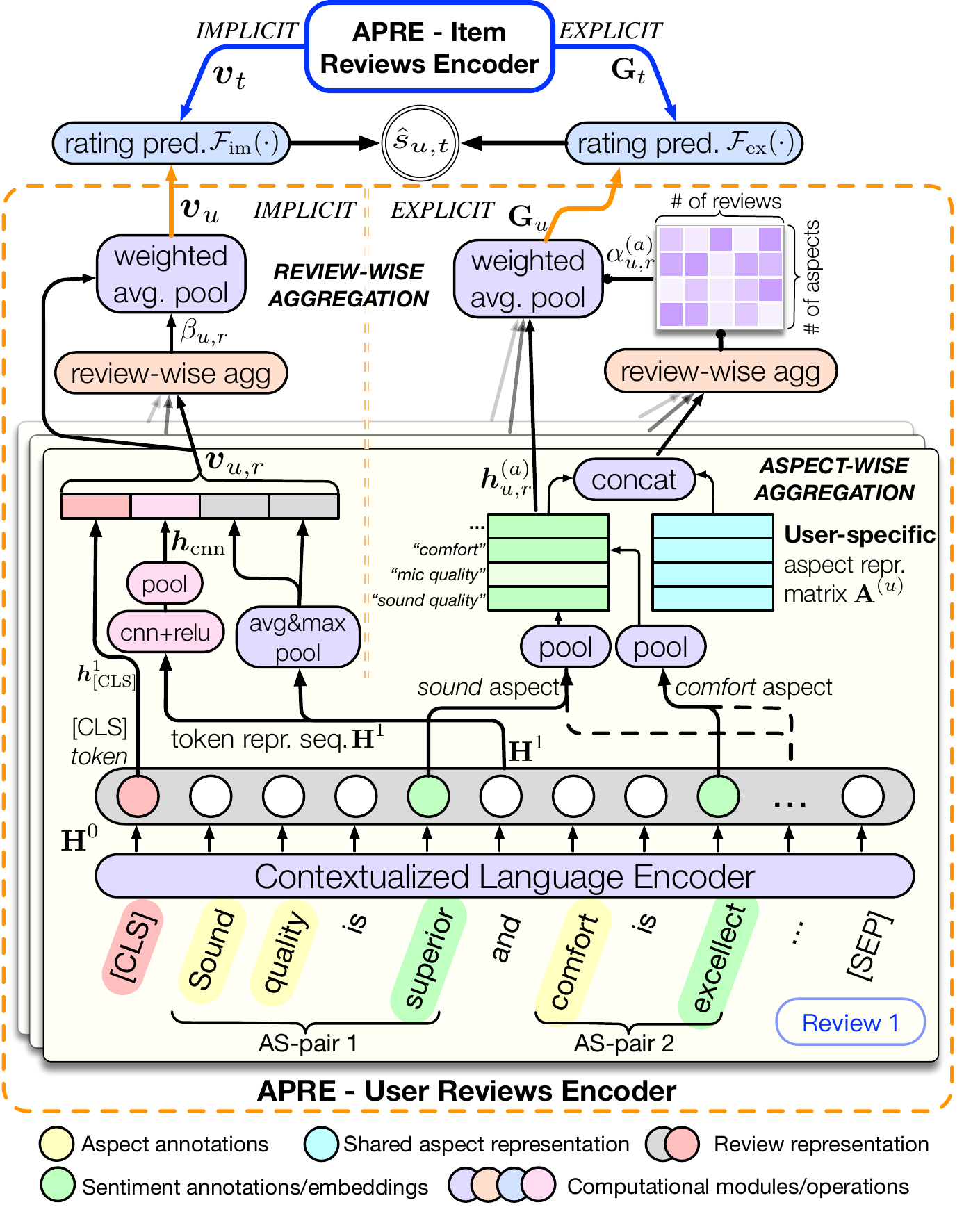}
    \caption{Pipeline of \rmn{} including a user review encoder in the orange dashed box and an item review encoder in the top blue box, each containing an implicit channel (left) and an aspect-based explicit channel (right). Internal details of item encoder are identical to the counterpart of user encoder and hence omitted.}
    \label{fig:part_two_pipeline}
\end{figure}

\rmn{}, depicted in Figure~\ref{fig:part_two_pipeline}, predicts ratings given reviews and the corresponding AS-pairs.
It first encodes language into embeddings, then learns explicit and implicit features, and finally computes the score regression.
One distinctive feature of \rmn{} is that it explicitly models the aspect information by incorporating a $d_a$-dimensional aspect representation $\veca_i\in\mathbb{R}^{d_a}$ in each side of the substructures for review encoding.
Let $\matA^{(u)}=\{\veca_1^{(u)}, \dots, \veca_k^{(u)}\}$ denotes the $k$ aspect embeddings for users and $\matA^{(t)}$ for items. $k$ is decided by the number of unique aspects in the AS-pair set.

\paragraph{Language encoding} The reviews are encoded into low-dimensional token embedding sequences by a fixed pre-trained BERT~\cite{bert}, a powerful transformer-based contextualized language encoder. 
For each review $r$
in $R^u$ or $R^t$, the resulting encoding $\matH^0\in\mathbb{R}^{(|r|+2)\times d_e}$ consists of $(|r|+2)$ $d_e$-dimensional contextualized vectors: $\matH^0=\{\vech^0_{\text{[CLS]}}, \vech^0_1, \dots, \vech^0_{|r|}, \vech^0_{\text{[SEP]}}\}$. \texttt{[CLS]} and \texttt{[SEP]} are two special tokens indicating starts and separators of sentences.
We use a trainable linear transformation, $\vech^1_i=\matW_\text{ad}^T\vech^0_i + \vecb_\text{ad}$, to adapt the BERT output representation $\matH^0$ to our task as $\matH^1$ where $\matW_\text{ad}\in \mathbb{R}^{d_e\times d_f}$, $\vecb_\text{ad}\in \mathbb{R}^{d_f}$, and $d_f$ is the transformed dimension of internal features.
BERT encodes the token semantics based upon the context which resolves the polysemy of certain sentiment terms, e.g., ``cheap'' is positive for \textit{price} but negative for \textit{quality}. This step transforms the sentiment encoding to attention-property modeling.

\paragraph{Explicit aspect-level attitude modeling} 
For aspect $a$ in the $k$ total aspects, we pull out all the contextualized representations of the sentiment words\footnote{BERT uses WordPiece tokenizer that can break an out-of-vocabulary word into shorter word pieces. If a sentiment word is broken into word pieces, we use the representation of the first word piece produced.} that modify $a$, and aggregate their representations to a single embedding of aspect $a$ in $r$ as 
$$
\vech^{(a)}_{u,r} = \sum\vech^1_j, w_j\in ST\cap r\text{ and } w_j\text{ modifies } a.$$
An observation by~\citet{chen2020aspect} suggests that users tend to use semantically consistent words for the same aspect in reviews. Therefore, sum-pooling can nicely handle both \textit{sentiments} and \textit{frequencies} of term mentions. 
Aspects that are not mentioned by $r$ will have $\vech^{(a)}_{u,r}=\boldsymbol{0}$.
To completely picture user $u$'s attentions to all aspects, we aggregate all reviews from $u$, i.e. $R^u$, using \textit{review-wise aggregation} weighted by $\alpha^{(a)}_{u,r}$ given in the equation below. $\alpha_{u,r}^{(a)}$ indicates the significance of each review's contribution to the overall understanding of $u$'s attention to aspect $a$
\begin{equation*}
     \alpha_{u,r}^{(a)}=\frac{\exp(\tanh(\vecw_\text{ex}^T [\vech_{u,r}^{(a)};\veca^{(u)}]))}{\sum_{r'\in R^u}\exp(\tanh(\vecw_\text{ex}^T [\vech_{u,r'}^{(a)};\veca^{(u)}]))},
     \label{eq:reviewwiseagg_ex}
\end{equation*}
where $[\cdot;\cdot]$ denotes the concatenation of tensors. $\vecw_\text{ex}\in\mathbb{R}^{(d_f+d_a)}$ is a trainable weight. With the usefulness distribution of $\alpha^{(a)}_{u,r}$, we aggregate the $\vech_{u,r}^{(a)}$ of $r\in R^u$ by weighted average pooling:
\begin{equation*}
\vecg_u^{(a)} = \sum_{r\in R^u} \alpha_{u,r}^{(a)} \vech_{u,r}^{(a)}.
\end{equation*}

Now we obtain the user attention representation for aspect $a$, $\vecg^{(a)}_u\in \mathbb{R}^{d_f}$. We use $\matG_u\in \mathbb{R}^{d_f\times k}$ to denote the matrix of $\vecg_u^{(a)}$. The item-tower architecture is omitted in Figure~\ref{fig:part_two_pipeline} since the \textbf{item property modeling} shares the identical computing procedure. It generates the item property representations $\vecg^{(a)}_t$ of $\matG_t$.
Mutual attention~\cite{liu2019daml,tay2018multi,dong2020asymmetrical} is \textit{not} utilized since the generation of user attention encodings $\matG_u$ is independent to the item properties and vice versa.

\paragraph{Implicit review representation}
It is acknowledged by existing works shown in Section~\ref{sec:relatedwork} that implicit semantic modeling is critical because some emotions are conveyed without explicit sentiment word mentions. For example, ``\textit{But this one feels like a pillow \dots}'' in R3 of Table~\ref{tab:review_examples} does not contain any sentiment tokens but expresses a strong satisfaction of the \textit{comfortableness}, which will be missed by the extractive annotation-based \tmn{}.

In \rmn{}, we combine a global feature $\vech^1_\text{[CLS]}$, a local context feature $\vech_\text{cnn}\in\mathbb{R}^{n_c}$ learned by a convolutional neural network (CNN) of output channel size $n_c$ and kernel size $n_k$ with max pooling, and two token-level features, average and max pooling of $\matH^1$ to build a comprehensive multi-granularity review representation $\vecv_{u,r}$:
\begin{align*}
\vecv_{u,r} &= \left[\vech^1_\text{[CLS]}; \vech_\text{cnn}; \text{\small{MaxPool}}(\matH^1); \text{\small{AvgPool}}(\matH^1)\right], \\
\vech_\text{cnn} &=\text{\small{MaxPool}}(\text{\small{ReLU}}(\text{\small{ConvNN\_1D}}(\matH^1))).
\end{align*}
We apply review-wise aggregation without aspects for latent review embedding $\vecv_u$
\begin{align*}
     \beta_{u,r}&=\frac{\exp(\tanh(\vecw_\text{im}^T \vecv_{u,r}))}{\sum_{r'\in R^u}\exp(\tanh(\vecw_\text{im}^T \vecv_{u,r'}))}, \\
     \vecv_u &= \sum\limits_{r\in R^u} \beta_{u,r} \vecv_{u,r}, 
\end{align*}
where $\beta_{u,r}$ is the counterpart of $\alpha^{(a)}_{u,r}$ in the implicit channel, $\vecw_\text{im}\in \mathbb{R}^{d_{im}}$ is a trainable parameter, and $d_{im}=3d_f+n_c$. Using similar steps, we can also obtain $\vecv_t$ for the item implicit embeddings.


\paragraph{Rating regression and optimization}
Implicit features $\vecv_u$ and $\vecv_t $ and explicit features $\matG_u$ and $\matG_t$ compose the input to the rating predictor to estimate the score $s_{u,t}$ by
\begin{equation*}
    \resizebox{\linewidth}{!}{%
    $\hat{s}_{u,t} = \underbrace{b_u + b_t}_{\text{biases}} + \underbrace{\mathcal{F}_\text{im}([\vecv_u; \vecv_t])}_{\text{implicit feature}} + \underbrace{\langle\vecgamma, \mathcal{F}_\text{ex}([\matG_u;\matG_t])\rangle}_{\text{explicit feature}}.$}
\end{equation*}
$\mathcal{F}_\text{im}: \mathbb{R}^{2d_{im}}\to \mathbb{R}$ and $\mathcal{F}_\text{ex}: \mathbb{R}^{2d_f\times k}\to \mathbb{R}^k$ are multi-layer fully-connected neural networks with ReLU activation and dropout to avoid overfitting. They model user attention and item property interactions in explicit and implicit channels, respectively. $\langle\cdot,\cdot\rangle$ denotes inner-product. $\vecgamma\in \mathbb{R}^{k}$ and $\{b_u,b_t\}\in \mathbb{R}$ are trainable parameters. The optimization function of the trainable parameter set $\Theta$ with an $L_2$ regularization weighted by $\lambda$ is  $$J(\Theta) = \sum_{r_{u,t}\in R}(s_{u,t} - \hat{s}_{u,t})^2 + L_2\text{-reg}(\lambda).$$  $J(\Theta)$ is optimized by back-propagation learning methods such as Adam~\cite{kingma2014adam}.
\section{Experiments}
\label{sec:experiments} 

\subsection{Experimental Setup} 
\paragraph{Datasets} We use \numdsl{} datasets from Amazon Review Datasets~\cite{he2016ups}\footnote{\url{https://jmcauley.ucsd.edu/data/amazon}} including AutoMotive (\textsf{AM}), Digital Music (\textsf{DM}), Musical Instruments (\textsf{MI}), Pet Supplies (\textsf{PS}), Sport and Outdoors (\textsf{SO}), Toys and Games (\textsf{TG}), and Tools and Home improvement (\textsf{TH}).
Their statistics are shown in Table~\ref{tab:data_stats}.
\begin{table*}[!ht]
    \centering
    \resizebox{0.9\linewidth}{!}{
   \setlength\tabcolsep{3pt}
    \begin{tabular}{rcccccccccc}
        \topline
        \rowcolor[gray]{.92}
            Dataset & Abbr. &\#Reviews & \#Users & \#Items & Density & Ttl. \#W & \#R/U & \#R/T &\#W/R  \\
        \midtopline
            AutoMotive & \textsf{AM} & 20,413 & 2,928 & 1,835 & 3.419$\times10^{-3}$ & 1.77M & 6.274 & 10.011 & 96.583 \\
            Digital Music & \textsf{DM} & 111,323 & 14,138 & 11,707 & 6.053$\times10^{-4}$ & 5.69M & 7.087 & 8.558 & 56.828 \\
            Musical Instruments & \textsf{MI} & 10,226 & 1,429 & 900 & \ul{7.156$\times10^{-3}$} & 0.96M & 6.440 & 10.226 & 103.958 \\
            Pet Supplies & \textsf{PS} & 157,376 & 19,854 & 8,510 & 8.383$\times10^{-4}$ & 14.23M & 7.134 & \ul{16.644} & 100.469 \\
            Sports and Outdoors & \textsf{SO} & \ul{295,434} &\ul{35,590} & \ul{18,357} & 4.070$\times10^{-4}$ & \ul{26.38M} & 7.471 & 14.484 & 99.199 \\
            Toys and Games & \textsf{TG} & 167,155 & 19,409 & 11,924 & 6.500$\times10^{-4}$ & 17.16M & \ul{7.751} & 12.616 & 114.047 \\
            Tools and Home improv. & \textsf{TH} & 134,129& 16,633 & 10,217 & 7.103$\times10^{-4}$ & 15.02M & 7.258 & 11.815 & \ul{124.429} \\
        \bottomrule
    \end{tabular}
    }
    \caption{The statistics of the \numdsl{} real-world datasets. (W: Words; U: Users; T: iTems; R: Reviews.)}
    \label{tab:data_stats}
\end{table*}


We use 8:1:1 as the train, validation, and test ratio for all experiments. 
Users and items with less than 5 reviews and reviews with less than 5 words are removed to reduce data sparsity. 

\paragraph{Baseline models}
\begin{table}[!h]
    \centering
    \resizebox{0.99\linewidth}{!}{
    \setlength\tabcolsep{3pt}
    \begin{tabular}{llccc}
    \topline
    \rowcolor[gray]{.92}
    Model & Reference & Cat. & U/T ID & Review \\
    \midtopline
    \textbf{MF} & - & Trad. &  \checkmark & \\
    \textbf{WRMF} & \citet{hu2008collaborative} & Trad.& \checkmark &\\
    \textbf{FM}& \citet{rendle2010factorization} & Trad. &\checkmark &   \\
    \textbf{ConvMF}& \citet{kim2016convolutional} & Deep& \checkmark & \checkmark \\
    \textbf{NeuMF} & \citet{he2017neural} & Deep & \checkmark &  \\
    \textbf{D-CNN}&\citet{zheng2017joint} & Deep & & \checkmark \\
    \textbf{D-Attn}&\citet{seo2017interpretable} & Deep & & \checkmark \\
    \textbf{NARRE}&\citet{chen2018neural} & Deep  & \checkmark & \checkmark  \\
    \textbf{ANR}&\citet{chin2018anr} & Deep  & & \checkmark \\
    \textbf{MPCN}&\citet{tay2018multi} & Deep  & \checkmark & \checkmark \\
    \textbf{DAML}&\citet{liu2019daml} & Deep & & \checkmark \\
    \textbf{AHN} &\citet{dong2020asymmetrical} & Deep & \checkmark & \checkmark \\
    \textbf{AHN-B} &Same as \textbf{AHN} & Deep & \checkmark & \checkmark \\
    \bottomrule
    \end{tabular}
    }
    \caption{Basics of compared baselines. Models' input is marked by ``\checkmark''. ``U'' and ``T'' denote Users and iTems. D-CNN represents DeepCoNN. AHN-B denotes the variant of AHN with BERT embeddings.}
    \label{tab:baselines}
\end{table}
\numbl{} baselines in \textit{traditional} and \textit{deep learning} categories are compared with the proposed framework. 
The pre-deep learning traditional approaches predict ratings solely based upon the entity IDs. 
Table~\ref{tab:baselines} introduces their basic profiles which are extended in Section~\ref{subsec:details_baselines}\bdm. Specially, \textbf{AHN-B} refers to AHN using pretrained BERT as the input embedding encoder. It is included to test the impact of the input encoders.
\paragraph{Evaluation metric}
We use Mean Square Error (MSE) for performance evaluation. Given a test set $R_\text{test}$, 
the MSE is defined by $$\text{MSE} = \frac{1}{|R_\text{test}|}\sum_{(u,r)\in R_\text{test}}(\hat{s}_{u,r} - s_{u,r})^2.$$

\paragraph{Reproducibility} We provide instructions to reproduce AS-pair extraction of \tmn{} and rating prediction of baselines and \rmn{} in Section~\ref{subsec:reproduce}\bdm. The source code of our models is publicly available on GitHub\footnote{\url{https://github.com/zyli93/ASPE-APRE}}.

\subsection{AS-pair Extraction of \tmn{}}
\label{subsec:aspair_extraction_results}
\begin{figure}[h]
\captionsetup{width=0.45\linewidth}
\centering
    \begin{minipage}[t]{.5\linewidth}
        \centering
        \includegraphics[width=\linewidth]{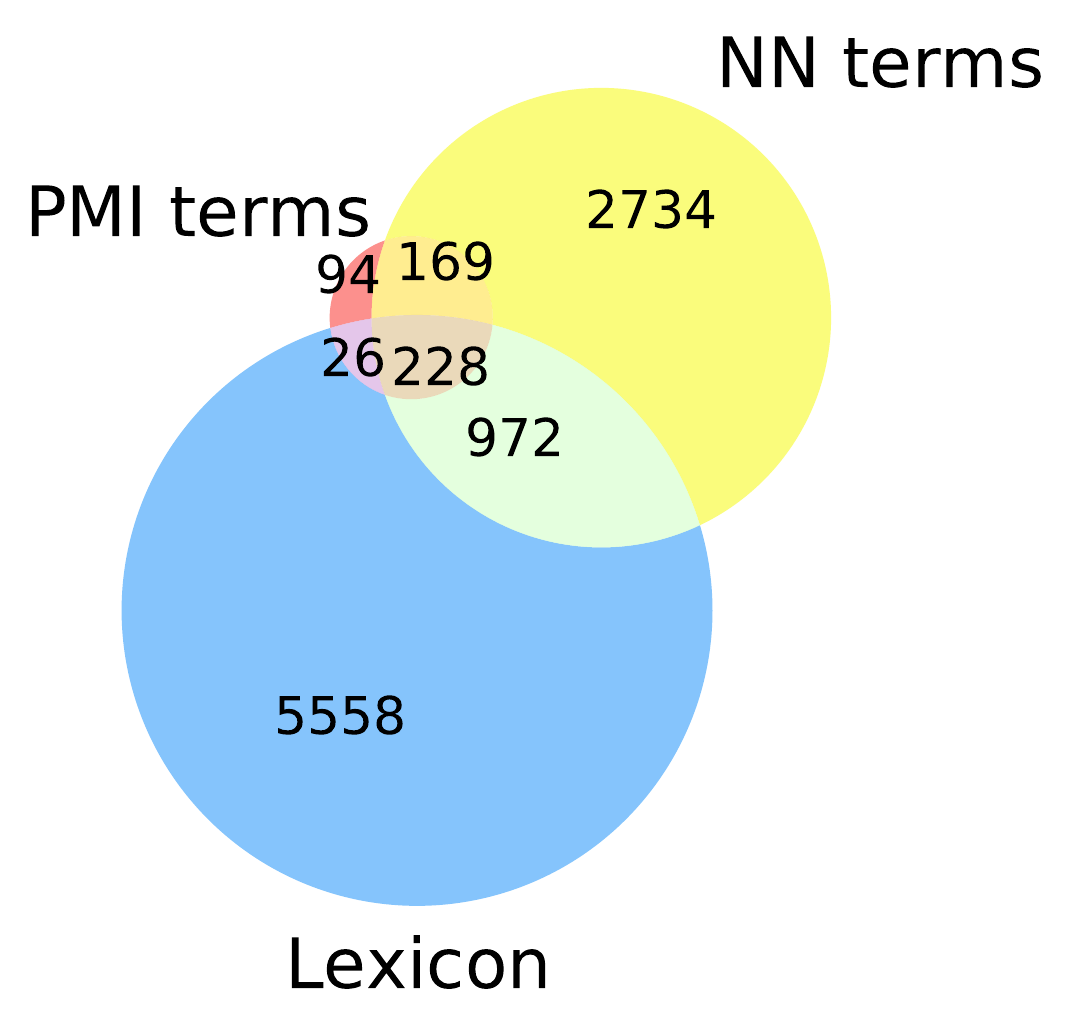}
        \captionof{figure}{Sources of sentiment terms from \textsf{AM}.}
        \label{fig:senti_term_venn}
        \vspace{-1em}
    \end{minipage}%
\hfill
    \begin{minipage}[t]{.5\linewidth}
        \centering
        \includegraphics[width=\linewidth]{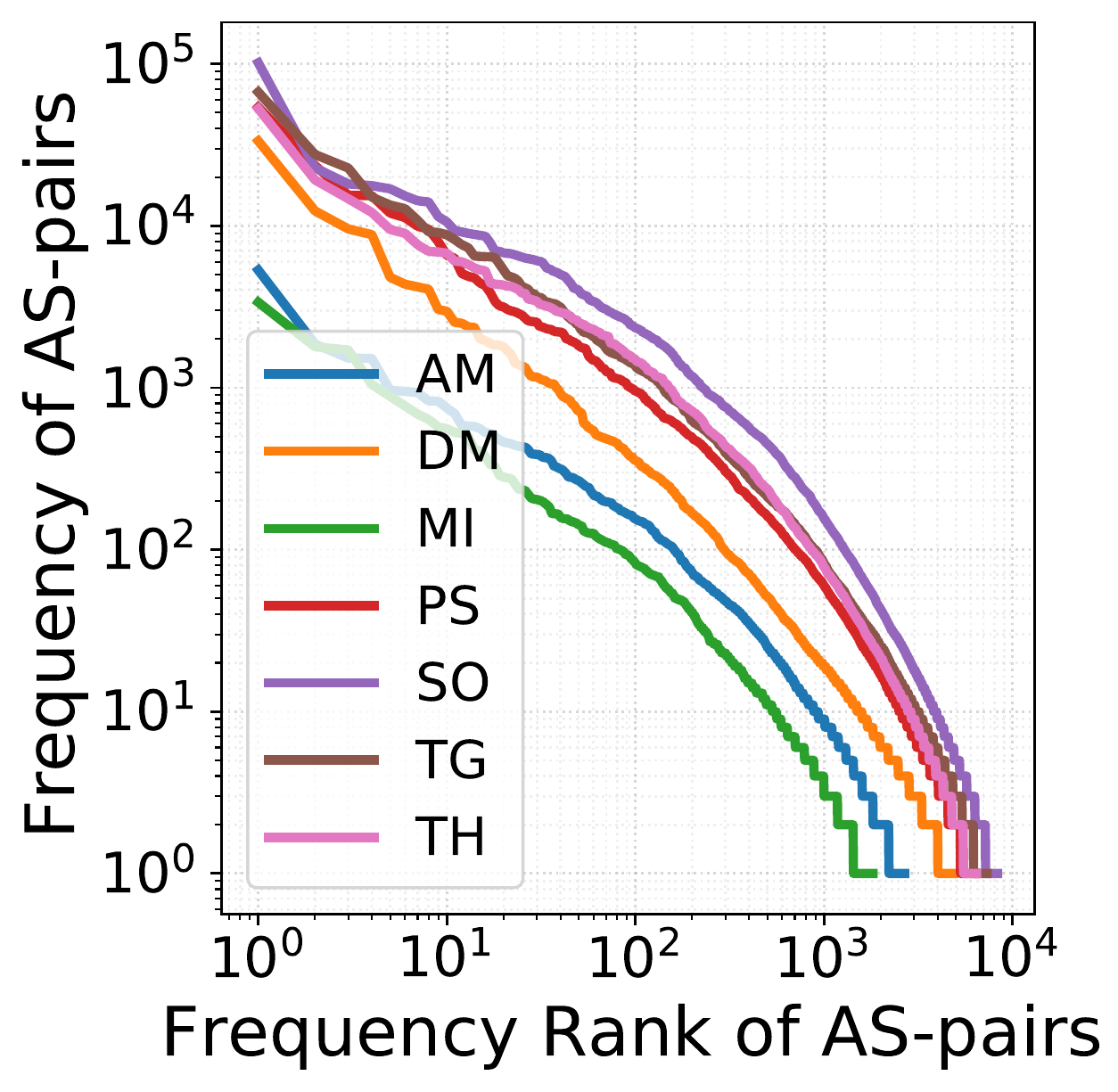}
        \captionof{figure}{Freq. rank vs. frequency of AS-pairs}
        \label{fig:aspair_dist_zipf_law}
        \vspace{-0.7em}
    \end{minipage}
\end{figure}

We present the extraction performance of unsupervised \tmn{}. The distributions of the frequencies of extracted AS-pairs in Figure~\ref{fig:aspair_dist_zipf_law} follow the trend of Zipf's Law with a deviation common to natural languages~\cite{li1992random}, meaning that \tmn{} performs consistently across domains. We show the qualitative results of term extraction separately. 

\paragraph{Sentiment terms} 
Generally, the AS-pair statistics given in Table~\ref{tab:aspair_stats}\bdm{} on different datasets are quantitatively consistent with the data statistics in Table~\ref{tab:data_stats}\bdm{} regardless of domain. Figure~\ref{fig:senti_term_venn} is a Venn diagram showing the sources of the sentiment terms extracted by \tmn{} from \textsf{AM}. All three methods are efficacious and contribute uniquely, which can also be verified by Table~\ref{tab:three_senti_terms}\bdm{} in Section~\ref{subsec:add_aspair_extraction}\bdm.

\paragraph{Aspect terms} Table~\ref{tab:top_aspect_terms} presents the most frequent aspect terms of all datasets. \textit{ItemTok} is ranked top as users tend to describe overall feelings about items. Domain-specific terms (e.g., \textit{car} in \textsf{AM}) and general terms (e.g., \textit{price}, \textit{quality}, and \textit{size}) are intermingled illustrating the comprehensive coverage and the high accuracy of the result of \tmn{}.

\begin{table}[!ht]
    \centering
        \resizebox{.99\linewidth}{!}{%
    \setlength\tabcolsep{1.5pt}
    \begin{tabular}{ccccccc}
        \topline
        \rowcolor[gray]{.92}
        \textsf{AM} & \textsf{DM} & \textsf{MI} & \textsf{PS} &  \textsf{SO} & \textsf{TG} & \textsf{TH}  \\
        \midtopline
         \textit{ItemTok}  & song    &  \textit{ItemTok} & \textit{ItemTok}  &  \textit{ItemTok}  & \textit{ItemTok}    & \textit{ItemTok}\\
         product           & \textit{ItemTok}  & sound & dog          &  knife        & toy  & light\\
         time              & album          & guitar   & food         &  quality      & game  & tool \\
         car               & music          & string   & cat          &  product     & piece & quality \\
         look              & time           &quality   & toy          & size       & quality & price\\
         price             & sound          & tone     & time         &  price      & child  & product \\
         quality           & voice          & price    & product      &  look      & color   & bulb\\
         light             & track          & pedal    & price        &  bag       & part    & battery\\
         oil               & lyric          & tuner    & treat        &  fit      & fun      & size\\
         battery           & version        & cable    & water        &  light       & size   & flashlight \\
        \bottomrule 
    \end{tabular}}
    \caption{High frequency aspects of the corpora.}
    \label{tab:top_aspect_terms}
\end{table}

\subsection{Rating Prediction of \rmn{}}
\label{subsec:evaluation_results}
\paragraph{Comparisons with baselines}
For the task of review-based rating prediction, a percentage increase \textbf{above $1\%$} in performance is considered significant~\cite{chin2018anr,tay2018multi}. According to Table~\ref{tab:mse}, our model outperforms all baseline models including the AHN-B on all datasets by a minimum of 1.337\% on \textsf{MI} and a maximum of 4.061\% on \textsf{TG}, which are significant improvements. 
It demonstrates (1) the superior capability of our model to make accurate rating predictions in different domains (Ours vs. the rest); (2) the performance improvement is NOT because of the use of BERT (Ours vs. AHN-B). 
AHN-B underperforms the original word2vec-based AHN\footnote{The authors of AHN also confirmed this observation.} because the weights of word2vec vectors are trainable while the BERT embeddings are fixed, which reduces the parameter capacity. Within baseline models, deep-learning-based models are generally stronger than entity ID-based traditional methods and recent ones tend to perform better. 

\paragraph{Ablation study} Ablation studies answer the question of which channel, explicit or implicit, contributes to the superior performance and to what extent? We measure their contributions by rows of \textit{w/o EX} and \textit{w/o IM} in Table~\ref{tab:mse}. \textit{w/o EX} presents the best MSEs of an \rmn{} variant \textit{without explicit} features under the default settings. The impact of AS-pairs is nullified. \textit{w/o IM}, in contrast, shows the best MSEs of an \rmn{} variant only leveraging the explicit channel while removing the implicit one (\textit{without implicit}).
We observe that the optimal performances of the single-channel variants all fall behind those of the dual-channel model, which reflects positive contributions from both channels. \textit{w/o IM} has lower MSEs than \textit{w/o EX} on several datasets showing that the explicit channel can supply comparatively more performance improvement than the implicit channel. It also suggests that the costly latent review encoding can be less effective than the aspect-sentiment level user and item profiling, which is a useful finding.

\newcommand{\ccg}{\cellcolor{green!10}} 
\newcommand{\ccy}{\cellcolor{yellow!10}} 
\newcommand{\ccc}{\cellcolor{cyan!10}} 
\newcommand{\ccgray}{\cellcolor{gray!10}} 
\newcommand{\ccr}{\cellcolor{red!10}} 

\hspace{-1em}
\begin{table}[!t]
    \centering
    \resizebox{1\linewidth}{!}{ 
    \setlength\tabcolsep{3pt}
    \begin{tabular}{rccccccc}
    \topline
    \rowcolor[gray]{.92}
    Models &  \textsf{AM} & \textsf{DM} & \textsf{MI} &\textsf{PS} & \textsf{SO} & \textsf{TG} & \textsf{TH} \\
    \midtopline
    \multicolumn{8}{c}{\small{\textsc{Traditional Models}}} \\
    \textbf{MF}   & 1.986   & 1.715 & 2.085 & 2.048 & 2.084 & 1.471 & 1.631 \\
    \textbf{WRMF} & 1.327   & 0.537 & 1.358 & 1.629 & 1.371 & 1.068 & 1.216 \\
    \textbf{FM}   &  1.082  & 0.436 & 1.146 & 1.458 & 1.212 & 0.922 & 1.050 \\
    \midrule
    \multicolumn{8}{c}{\small{\textsc{Deep Learning-based Models}}} \\
    \textbf{ConvMF} & 1.046      & 0.407      & 1.075      & 1.458      & 1.026      & 0.986      & 1.104      \\
    \textbf{NeuMF}  & 0.901      & 0.396      & 0.903      & 1.294      & 0.893      & 0.841      & 1.072      \\
    \textbf{D-Attn} & 0.816      & 0.403      & 0.835      & 1.264      & 0.897      & 0.887      & 0.980      \\
    \textbf{D-CNN}  & 0.809      & 0.390      & 0.861      & 1.250      & 0.894      & 0.835      & 0.975      \\
    \textbf{NARRE}  & 0.826      & 0.374      & 0.837      & 1.425      & 0.990      & 0.908      & \ul{0.958} \\
    \textbf{MPCN}   & 0.815      & 0.447      & 0.842      & 1.300      & 0.929      & 0.898      & 0.969      \\
    \textbf{ANR}    & 0.806      & 0.381      & 0.845      & 1.327      & 0.906      & 0.844      & 0.981      \\
    \textbf{DAML}   & 0.829      & \ul{0.372} & 0.837      & \ul{1.247} & 0.893      & \ul{0.820} & 0.962      \\
    \textbf{AHN-B}  & 0.810      & 0.385      & 0.840      & 1.270      & 0.896      & 0.829      & 0.976   \\
    \textbf{AHN}    & \ul{0.802} & 0.376      & \ul{0.834} & 1.252      & \ul{0.887} & 0.822      & 0.967      \\
    \midrule
    \rowcolor{green!10}
    \multicolumn{8}{c}{\small{\textsc{Our Models and Percentage Improvements}}} \\
    \rowcolor{green!10}
    \textbf{Ours} & \textbf{0.791} & \textbf{0.359} & \textbf{0.823} & \textbf{1.218} & \textbf{0.863} & \textbf{0.788} & \textbf{0.936}\\ 
    \rowcolor{green!10}
    $\Delta(\%)$   & 1.390  & 3.621 &1.337 & 2.381 & 2.784 & 4.061 & 2.350 \\
    \rowcolor{green!10}
    \textbf{Val.}   & 0.790 & 0.362 & 0.821      & 1.216 & 0.860 & 0.790      & 0.933 \\
    \midrule
    \rowcolor{yellow!10}
    \multicolumn{8}{c}{\small{\textsc{Ablation Studies}}} \\
    \rowcolor{yellow!10}
    \textbf{w/o EX} & 0.814      & 0.379      & 0.833 & 1.244.     & 0.882      & 0.796 & 0.965 \\
    \rowcolor{yellow!10}
    \textbf{w/o IM} & 0.798 & 0.374 & 0.863      & 1.226 & 0.873 & 0.798      & 0.956 \\
    \bottomrule
    \end{tabular}}
    \caption{MSE of baselines, our model (\textbf{Ours} for test and \textbf{Val.} for validation), and variants. The row of $\Delta$ calculates the percentage improvements over the \ul{best baselines}. All reported improvements over the best baselines are statistically significant with $p$-value $<$ 0.01.}
    \label{tab:mse}
\end{table}

              

\begin{figure}[t!]
    \centering
    \begin{subfigure}[t]{0.5\linewidth}
    \captionsetup{width=0.95\linewidth}
        \centering
        \includegraphics[width=0.98\linewidth]{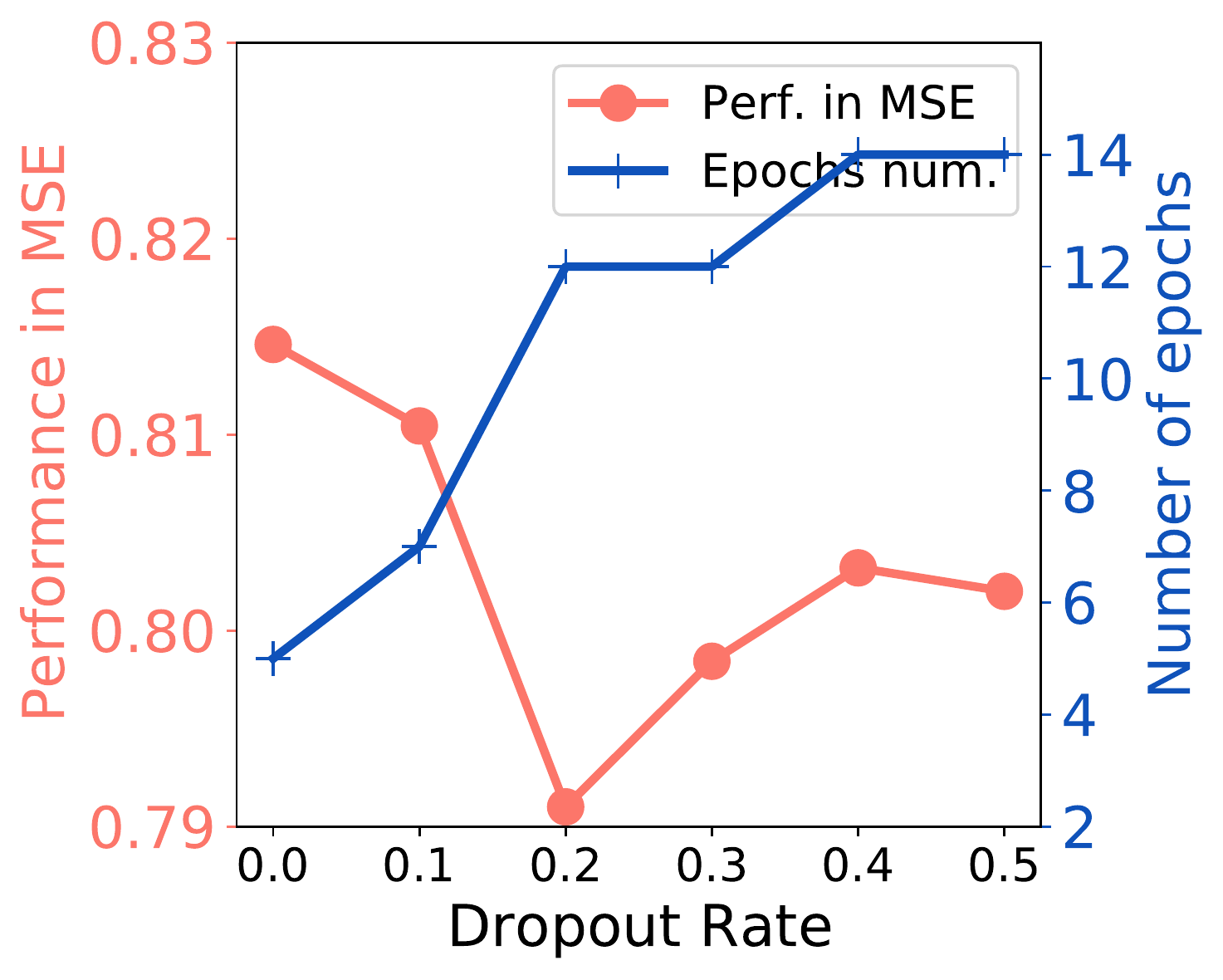}
        \caption{Dropout vs. MSE and Ep.}
    \end{subfigure}%
    ~ 
    \begin{subfigure}[t]{0.5\linewidth}
    \captionsetup{width=0.95\linewidth}
        \centering
        \includegraphics[width=0.98\linewidth]{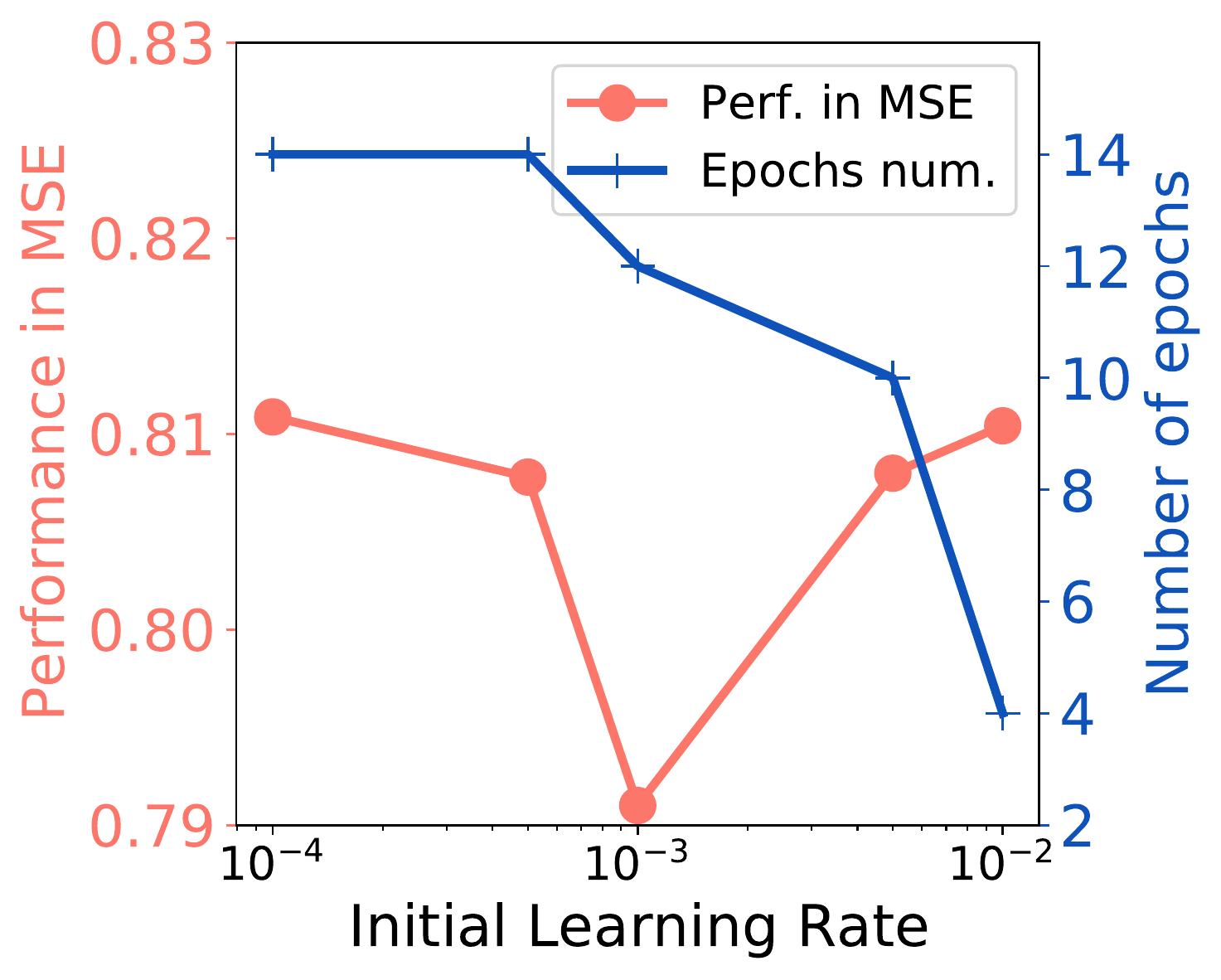}
        \caption{Init. LR vs. MSE and Ep.}
    \end{subfigure}
    \caption{Hyper-parameter searching and sensitivity}
    \label{fig:parameter_sensitivity}
\end{figure}

\paragraph{Hyper-parameter sensitivity}
A number of hyper-parameter settings are of interest, e.g., dropout, learning rate (LR), internal feature dimensions ($d_a$, $d_f$, $n_c$, and $n_k$), and regularization weight $\lambda$ of the $L_2$-reg in $J(\Theta)$. We run each set of experiments on sensitivity search 10 times and report the average performances. We tune dropout rate in $[0,0.1,0.2,0.3,0.4,0.5]$ and LR\footnote{The reported LRs are initial since Adam and a LR scheduler adjust it dynamically along the training.} in $[0.0001, 0.0005, 0.001, 0.005, 0.01]$
with other hyper-parameters set to default, and report in Figure~\ref{fig:parameter_sensitivity} the minimum MSEs and the epoch numbers (Ep.) on \textsf{AM}. For dropout, we find the balance of its effects on avoiding overfitting and reducing active parameters at 0.2. Larger dropouts need more training epochs. For LR, we also target a balance between training instability of large LRs and overfitting concern of small LRs, thus 0.001 is selected. Larger LRs plateau earlier with fewer epochs while smaller LRs later with more. Figure~\ref{fig:add_param_sensitivity}\bdm{} analyzes hyper-parameter sensitivities  to changes on internal feature dimensions ($d_a$, $d_f$, and $n_c$), CNN kernel size $n_k$, and $\lambda$ of $L_2$-reg weight. 

\paragraph{Efficiency} 
A brief run time analysis of \rmn{} is given in Table~\ref{tab:runtime}.
The model can run fast with all data in GPU memory such as \textsf{AM} and \textsf{MI}, which demonstrates the efficiency of our model and the room for improvement on the run time of datasets that cannot fit in the GPU memory.
The efficiency of \tmn{} is less critical since it only runs once for each dataset.
\begin{table}[]
    \centering
    \setlength\tabcolsep{3pt}
    \resizebox{0.9\linewidth}{!}{
    \begin{tabular}{ccccccc}
    \topline
    \rowcolor[gray]{.92}
    \textsf{AM} & \textsf{DM} & \textsf{MI} &\textsf{PS} & \textsf{SO} & \textsf{TG} & \textsf{TH} \\
    \midtopline
     127s$^*$ & 31min & 90s$^*$ & 36min & 90min & 51min & 35min \\
    \bottomrule
    \end{tabular}}
    \caption{Per epoch run time of \rmn{} on the seven datasets. The run time of \textsf{AM} and \textsf{MI}, denoted by ``$^*$'', is disproportional to their sizes since they can fit into the GPU memory for acceleration.}
    \label{tab:runtime}
\end{table}

\subsection{Case Study for Interpretation}
\label{sec:case_study}
Finally, we showcase an interpretation procedure of the rating estimation for an instance in \textsf{AM}: how does \rmn{} predict $u_*$'s rating for a smart driving assistant $t_*$ using the output AS-pairs of \tmn{}?
We select seven example aspect categories with all review snippets mentioning those categories. Each category is a set of similar aspect terms, e.g., \{\textit{look}, \textit{design}\} and \{\textit{beep}, \textit{sound}\}. Without loss of generality, we refer to the categories as aspects.
Table~\ref{tab:case_study_reviews} presents the aspects and review snippets given by $u_*$ and received by $t_*$ with AS-pairs annotations. Three aspects, \{\textit{battery}, \textit{install}, \textit{look}\}, are shared (yellow rows). Each side has two unique aspects never mentioned by the reviews of the other side: \{\textit{materials}, \textit{smell}\} of $u_*$ (green rows) and \{\textit{price}, \textit{sound}\} of $t_*$ (blue rows).

\rmn{} measures the aspect-level contributions of user-attention and item-property interactions by the last term of $s_{u,t}$ prediction, i.e., $\langle\vecgamma,\mathcal{F}_\text{ex}([\matG_u;\matG_t])\rangle$. The contribution on the $i$th aspect is calculated by the $i$th dimension of $\gamma$ times the $i$th value of $\mathcal{F}_\text{ex}([\matG_u;\matG_t])$ which is shown in Table~\ref{tab:case_study_summary}. The top two rows summarize the attentions of $u_*$ and the properties of $t_*$. \textit{Inferred Impact} states the interactional effects of user attentions and item properties based on our assumption that attended aspects bear stronger impacts to the final prediction.
On the overlapping aspects, the inferior property of \textit{battery} produces the only negative score (-0.008) whereas the advantages on \textit{install} and \textit{look} create positive scores (0.019 and 0.015), which is consistent with the inferred impact. Other aspects, either \textit{unknown} to user attentions or to item properties, contribute relatively less: $t_*$'s unappealing \textit{price} accounts for the small score 0.009 and the mixture property of \textit{sound} accounts for the 0.006.

This case study demonstrates the usefulness of the numbers that add up to $\hat{s}_{u,t}$. Although small in scale, they carry significant information of valued or disliked aspects in $u_*$'s perception of $t_*$. This process of decomposition is a great way to interpret model prediction on an aspect-level granularity, which is a capacity that other baseline models do not enjoy.

\newcommand{\asp}[1]{\textit{\ul{#1}}}
\newcommand{\senti}[1]{\textit{\textbf{#1}}}
\newcommand{\supp}[1]{\textit{#1}}
\newcommand{\xmark}{\textcolor{red}{\ding{55}}}  
\newcommand{\gcmark}{\mygreen{\ding{51}}}
\newcommand{\bcmark}{\textcolor{blue}{\ding{51}}}
\newcommand{\revUserBattery}{\small{\textbf{[To $t_1$]} After leaving this attached to my car for two days of non-use I have a \senti{dead} \asp{battery}. Never had a \senti{dead} \asp{battery} \dots, so \supp{I am blaming this device.}}}
\newcommand{\revUserInstall}{\small{\textbf{[To $t_2$]} \asp{This} was \senti{unbelievably easy} to install. I have done \dots. The real key \dots the \asp{installation} is so \senti{easy}. \textbf{[To $t_3$]} There were \senti{many} \asp{installation options}, but once \dots, \supp{they clicked on easily.}}}
\newcommand{\revUserLook}{\small{\textbf{[To $t_3$]} It was \supp{not perfect and not shiny}, but it did \asp{look} \senti{better}. \textbf{[To $t_4$]} It takes some \senti{elbow} \asp{grease}, but the \supp{results are remarkable.}}}

\newcommand{\revUserMaterial}{\small{\textbf{[To $t_5$]} \asp{The plastic} however is \senti{very thin} and \asp{the cap} is \senti{pretty cheap}. \textbf{[To $t_6$]} Great value. \dots. They are \senti{very hard} \asp{plastic}, so they don't mark up panels.}}
\newcommand{\revUserSmell}{\small{\textbf{[To $t_7$]} This has a \senti{terrible} \asp{smell} that really lingers awhile. It goes on green. \dots}}

\newcommand{\revItemBattery}{\small{\textbf{[From $u_1$]} The reason this won't work on an iPhone 4 or \dots because it uses \senti{low} \asp{power} Bluetooth, \dots. (\xmark)}}
\newcommand{\revItemInstall}{\small{\textbf{[From $u_2$]} Your mileage and gas mileage and cost of fuel is tabulated for each trip- \asp{Installation} is \senti{pretty simple} - but it \dots. (\gcmark)}}
\newcommand{\revItemLook}{\small{\textbf{[From $u_3$]} Driving habits, fuel efficiency, and engine health are \senti{nice} \asp{features}. The \asp{overall design} is \senti{nice} and \supp{easy to navigate}. (\gcmark)}}
\newcommand{\revItemPrice}{\small{\textbf{[From $u_4$]} In fact, there are similar products to this available at a \senti{much lower} \asp{price} that do work with \dots (\xmark) }}
\newcommand{\revItemSound}{\small{\textbf{[From $u_5$]} The Link device makes an \senti{audible} \asp{sound} when you go over 70 mpg, brake hard, or accelerate too fast. (\gcmark) \textbf{[From $u_6$]} Also, the \asp{beep} the link device makes \dots sounds \senti{really cheapy}. (\xmark)}}

\newcolumntype{Q}{>{\raggedright\arraybackslash}p{0.92\linewidth}}

\begin{table}[!h]
    \centering
    \resizebox{0.99\linewidth}{!}{
    \begin{tabular}{cQ}
    \toprule
    \multicolumn{2}{c}{\textit{From reviews given by user} $u_*$. \textit{All aspects attended} (\bcmark).} \\
    \midrule
    \rowcolor{yellow!10} battery & \revUserBattery \\
    \midrule
    \rowcolor{yellow!10} install & \revUserInstall \\
    \midrule
    \rowcolor{yellow!10} look    & \revUserLook \\
    \midrule
    \rowcolor{green!10} material & \revUserMaterial \\
    \midrule 
    \rowcolor{green!10} smell & \revUserSmell \\
    \midrule
    \midrule
    \multicolumn{2}{c}{\textit{From reviews received by item} $t_*$.} \\
    \midrule
    \rowcolor{yellow!10} battery & \revItemBattery\\
    \midrule
    \rowcolor{yellow!10} install & \revItemInstall \\
    \midrule
    \rowcolor{yellow!10} look    & \revItemLook \\
    \midrule
    \rowcolor{cyan!10} price & \revItemPrice \\
    \midrule 
    \rowcolor{cyan!10} sound & \revItemSound \\
    \bottomrule
    \end{tabular}}
    \caption{Examples of reviews given by $u_*$ and received by $t_*$ with \asp{Aspect}-\senti{Sentiment} pair mentions as well as \supp{other sentiment evidences} on seven example aspects.}
    \label{tab:case_study_reviews}
\end{table}

\newcommand{\negImp}{\textcolor{red}{\textit{\textbf{Neg.}}}}
\newcommand{\posImp}{\mygreen{\textbf{\textit{Pos.}}}}
\newcommand{\unc}{\textcolor{gray!90}{\textit{Unk.}}}

\begin{table}[!h]
    \centering
     \resizebox{\linewidth}{!}{
     \setlength\tabcolsep{2pt}
    \begin{tabular}{cccccccc}
        \topline
        \rowcolor[gray]{0.92}
         Aspects & material & smell & battery & install & look & price & sound \\
        \midtopline
        Attn. of $u_*$ & \ccg \bcmark & \ccg \bcmark & \ccy \bcmark & \ccy \bcmark & \ccy \bcmark & n/a    & n/a \\
        Prop. of $t_*$ &  n/a   & n/a    & \ccy \xmark & \ccy \gcmark & \ccy \gcmark & \ccc \xmark & \ccc \gcmark/\xmark \\
        Inferred Impact & \unc & \unc & \negImp & \posImp & \posImp & \unc & \unc  \\
        \midrule
        $\vecgamma_i\mathcal{F}_{\text{ex}} (\cdot)_i$ ($\times 10^{-2}$) & \ccgray 1.0 & \ccgray 0.8 & \ccr -0.8 & \ccg 1.9 & \ccg 1.5 & \ccgray  0.9 & \ccgray  0.6 \\ 
        \bottomrule
    \end{tabular}}
    \caption{Attentions and properties summaries, inferred impacts, and the learned aspect-level contributions.}
    \label{tab:case_study_summary}
\end{table}

In Section~\ref{subsec:case_study_ii}\bdm{}, another case study indicates that a certain imperfect item property without user attentions only inconsiderably affects the rating although the aspect is mentioned by the user's reviews.
\section{Conclusion}
\label{sec:conclusion}
In this work, we propose a tightly coupled two-stage review-based rating predictor, consisting of an Aspect-Sentiment Pair Extractor (\tmn{}) and an Attention-Property-aware Rating Estimator (\rmn{}). 
\tmn{} extracts aspect-sentiment pairs (AS-pairs) from reviews and \rmn{} learns explicit user attentions and item properties as well as implicit sentence semantics to predict the rating.
Extensive quantitative and qualitative experimental results demonstrate that \tmn{} accurately and comprehensively extracts AS-pairs without using domain-specific training data and \rmn{} outperforms the state-of-the-art recommender frameworks and explains the prediction results taking advantage of the extracted AS-pairs.

Several challenges are left open such as fully or weakly supervised open domain AS-pair extraction and end-to-end design for AS-pair extraction and rating prediction. We leave these problems for future work.

\section*{Acknowledgement}
We would like to thank the reviewers for their helpful comments. 
The work was partially supported by NSF DGE-1829071 and NSF IIS-2106859.

\section*{Broader Impact Statement}

This paper proposes a rating prediction model that has a great potential to be widely applied to recommender systems with reviews due to its high accuracy. In the meantime, it tries to relieve the unjustifiability issue for black-box neural networks by suggesting what aspects of an item a user may feel satisfied or dissatisfied with. The recommender system can better understand the rationale behind users' reviews so that the merits of items can be carried forward while the defects can be fixed. As far as we are concerned, this work is the first work that takes care of both rating prediction and rationale understanding utilizing NLP techniques.

We then address the generalizability and deployment issues. Reported experiments are conducted on different domains in English with distinct review styles and diverse user populations. We can observe that our model performs consistently which supports its generalizability.
Ranging from smaller datasets to larger datasets, we have not noticed any potential deployment issues. Instead, we notice that stronger computational resources can greatly speed up the training and inference and scale up the problem size while keeping the major execution pipeline unchanged. 

In terms of the potential harms and misuses, we believe they and their consequences involve two perspectives: (1) the harm of generating inaccurate or suboptimal results from this recommender; (2) the risk of misuse (attack) of this model to reveal user identity. For point (1), the potential risk of suboptimal results has little impact on the major function of online shopping websites since recommenders are only in charge of suggestive content. For point (2), our model does not involve user and item ID modeling. Also, we aggregate the user reviews in the representation space so that user identity is hard to infer through reverse-engineering attacks.
In all, we believe our model has little risk of causing dysfunction of online shopping platforms and leakages of user identities.

\bibliography{acl2021}
\bibliographystyle{acl_natbib}

\clearpage
\appendix

\section{Supplementary Materials}
\label{sec:appendix}
\subsection{Introduction}
This document is the Supplementary Materials for \textit{\myTitle}.
It contains supporting materials that are important but unable to be completely covered in the main transcript due to the page limits. 

\subsection{Methods}
\subsubsection{Pseudocode of \tmn{}}
\label{subsubsec:pseudocode}
Although Section~\ref{subsec:extractor} is self-explanatory, we would like to explain the AS-pair generation process in Section~\ref{subsec:aspair_generation} in detail by Algorithm~\ref{alg:pseudocode_aspair_building}. It leverages the sentiment term set $ST$ obtained from Section~\ref{subsec:senti_term_extraction}, a dependency parser, and WordNet synsets to build AS-pairs.
\newcommand\mycommfont[1]{\footnotesize\ttfamily\textcolor{blue}{#1}}
\SetArgSty{textnormal}
\SetCommentSty{mycommfont}
\SetKw{And}{and}
\SetKw{Or}{or}
\SetKw{Ret}{return}
\SetKwInput{KwInput}{Input}                
\SetKwInput{KwOutput}{Output}              
  
\begin{algorithm}[h]
\DontPrintSemicolon
  \KwInput{Sentiment terms $ST$, dependency parser \texttt{DepParser}, threshold $c$.}
  \KwOutput{AS-pairs}
  \KwData{Review-rating corpus $R$; WordNet with \texttt{synsets}.}
  \tcc{Initialize AS-pair candidate and AS-pair sets}
  \textit{AS-cand}, \textit{AS-pairs} $\xleftarrow{} \emptyset$, $\emptyset$ \\
  \tcc{Extract AS-pair candidates.}
  \ForEach{review $r\in R$}{
    dep-graph$_r \xleftarrow{}$\texttt{DepParser}$(r)$ \\
    \ForEach{dependency relation $r_\text{dep}$ in dep-graph$_r$}
    {
        \If{$r_\text{dep}$ is \texttt{nsubj+acomp} \Or $r_\text{dep}$ is \texttt{amod}}{
            Add corresponding (\textit{noun, adj.}) tuple to \textit{AS-cand} (Figure~\ref{tab:dep_rules})
        }
    }
  }
  \tcc{Merge synonym aspects}
  \ForEach{(\textit{noun,adj.}) tuple $\in$ \textit{AS-cand}}{
    MergeSynAspect(\texttt{synsets}, \textit{noun})
  }
  \tcc{Filter out non-AS-pairs by $ST$ and frequency threshold $c$.}
  \ForEach{(\textit{noun, adj.}) tuple $\in$ \textit{AS-cand}}{
    \If{\textit{adj.}$\in ST$ \And Freq$[$\textit{noun}$]> c$}{
        Add (\textit{noun, adj.}) to \textit{AS-pairs}
    }
  }
\Ret \textit{AS-pairs}
\caption{AS-pairs Generation}
\label{alg:pseudocode_aspair_building}
\end{algorithm}

\subsection{Experiments}
This section exhibits additional content regarding the experiments such as a detailed experimental setup, the instructions to reproduce the baselines and our model, supplemental experimental results, and another case study.
We hope the critical content help readers gain deeper insight into the performance of the proposed framework.


\subsubsection{Reproducibility of \tmn{} and \rmn{}}
\label{subsec:reproduce}
\tmn{}+\rmn{} is implemented in Python (3.6.8) with PyTorch (1.5.0) and run with a single 12GB Nvidia Titan Xp GPU.
The code is available on GitHub\footnote{\url{https://github.com/zyli93/ASPE-APRE}} and comprehensive instructions on how to reproduce our model are also provided. The default hyper-parameter settings for the results in Section~\ref{subsec:evaluation_results} are as follows: 
\paragraph{\tmn{}} In the AS-pair extraction stage, we set the size of $ctx$ to 5 and the PMI term quota $q$ to 400 for both polarities. The counting thresholds $c$ for different datasets are given in Table~\ref{tab:aspair_stats}. SDRN~\cite{chen2020synchronous} utilized for term extraction is trained under the default settings in the source code\footnote{\url{https://github.com/chenshaowei57/SDRN}} with the SemEval 14/15 datasets mentioned in Section~\ref{sec:relatedwork}. \texttt{spaCy}\footnote{\url{https://spacy.io}}, a Python package specialized in NLP algorithms, provides the dependency parsing pipeline. 
\paragraph{\rmn{}} In the rating prediction stage, we use a pre-trained BERT model with 4 layers, 4 heads, and 256 hidden dimensions (``BERT-mini'') for manageable GPU memory consumption. The BERT parameters (or weights) are fixed. The BERT tokenizer and model are loaded from the Hugging Face model repository\footnote{\url{https://huggingface.co/google/bert_uncased_L-4_H-256_A-4}}. The initial learning rate is set to \num{0.001} with two adjusting mechanisms: (1) the Adam optimizer $(\beta_1,\beta_2)=(0.9, 0.999)$ (the default setting in PyTorch); (2) a learning rate scheduler, \texttt{StepLR}, with step size as 3 and \texttt{gamma} as 0.8. Dropout is set to 0.2 for both towers. $d_f$, $d_a$, and $n_c$ are all set to 200 for consistency. The CNN kernel size is 4. The $L_2$-reg weight, $\lambda$, is set globally to 0.0001. We use a clamp function to constrain the predictions in the interval $(1.0, 5.0)$. 


\subsubsection{\tmn{}: Additional Experimental Results of AS-pair Extraction}
\label{subsec:add_aspair_extraction}
We present in Table~\ref{tab:aspair_stats} the statistics of the extracted AS-pairs of the corpora which are quantitatively consistent with the data statistics in Table~\ref{tab:data_stats} regardless of domain.

\begin{table}[!h]
    \centering
    \resizebox{\linewidth}{!}{
    \setlength\tabcolsep{3pt}
    \begin{tabular}{ccccccc}
    \topline
    \rowcolor[gray]{.92}
    Data & $c$ & \#AS-pairs/R &  \#A/U & \#A/T & \#A & \#S \\
    \midtopline
    \textsf{AM} & 50  &  3.076 & 12.681 & 16.284 & 291 & 8,572  \\
    \textsf{DM} & 100  &  1.973 & 5.792 & 8.380 & 296  & 9,781   \\
    \textsf{MI} & 50  &  3.358 & 12.521 & 16.323 & 167 & 8,143\\
    \textsf{PS} & 150  &  3.445 & 14.886 & 23.893 & 529 & 12,563 \\
    \textsf{SO} & 250  &  4.078 & 19.401 & 28.314 & 747 & 17,195 \\
    \textsf{TG} & 150  &  4.482 & 19.053 & 26.657 & 680 & 13,972 \\
    \textsf{TH} & 150  &  5.235 & 22.833 & 29.816 & 659 & 14,145\\
    \bottomrule
    \end{tabular}}
    \caption{Statistics of unsupervised AS-pair extraction. $c$: frequency threshold; R: reviews; U: users; T: items.}
    \label{tab:aspair_stats}
\end{table}

We provide Table~\ref{tab:three_senti_terms} ancillary to the Venn diagram in Figure~\ref{fig:senti_term_venn} and the corresponding conclusion in Section~\ref{subsec:aspair_extraction_results}.
Table~\ref{tab:three_senti_terms} illustrates the contributions of the three distinct sentiment term extraction methods discussed in Section~\ref{subsec:extractor}, namely PMI-based method, neural network-based method, and lexicon-based method. All three methods can extract useful sentiment-carrying words in the domain of Automotive. Their contributions cannot overwhelm each other, which strongly explains the necessity of the unsupervised methods for term extraction in the domain-general usage scenario. Altogether they provide comprehensive coverage of sentiment terms in \textsf{AM}.
\begin{table*}[!ht]
    \centering
    \resizebox{0.9\linewidth}{!}{%
    \begin{tabular}{ccccccc}
    \topline
    \rowcolor[gray]{.92}

     $P$ only & $N$ only & $L$ only & $P\cap N\backslash L$  & $P\cap L\backslash N$ & $N\cap L \backslash P$ & $P\cap N\cap L$ \\
    \midtopline 
    countless & therapeutic & fateful & ultimate & uplifting & dazzling & amazing \\
    dreamy & vital  & poorest & new & concerned & costly & beautiful \\
    edgy & uncanny & tedious & rhythmic & joyful & devastated & classic  \\
    entire & adept & unwell & generic & bombastic & faster & delightful  \\
    forgettable & fulfilling & joyous & atmospheric & unforgettable & graceful & enjoyable \\
    melodious & attracted & illegal & greater & phenomenal & affordable & fantastic   \\
    moral &  celestial & noxious  & supernatural & inventive & supreme & gorgeous  \\
    propulsive & harmonic & lovable & contemporary & classy & robust & horrible   \\
    tasteful & newest & crappy & surprising & insightful & useless & inexpensive   \\
    uninspired & enduring & arduous & tremendous & masterful & unpredictable & magnificent  \\
    \bottomrule 
    \end{tabular}}
    \caption{Example sentiment terms of each part of the Venn diagram (Figure~\ref{fig:senti_term_venn}) from \textsf{AM} dataset. We use $P$ (PMI), $N$ (Neural network), and $L$ (Lexicon) to denote the produced sentiment term sets of the three methods, respectively. Operator $\backslash$ denotes set minus, e.g., $P\cap L \backslash N$ refers to the set of terms that are in both $P$ and $L$ but not in $N$. All sets contain commonly-used sentimental adjectives that can modify automotive items. This figure strongly explains why three methods are all necessary for term extraction in non-domain-specific use cases. They all have unique contributions to the sentiment term set for larger coverage.}
    \label{tab:three_senti_terms}
\end{table*}

\subsubsection{\rmn{}: Information of Baselines}
\label{subsec:details_baselines}
We introduce baseline models mentioned in Table~\ref{tab:baselines} including the source code of the software and the key parameter settings.
For the fairness of comparison, we only compare the models that have \textbf{open-source} implementations.
\paragraph{MF, WRMF, FM, and NeuMF\footnote{Source code of MF, WRMF, FM, and NeuMF is available in \texttt{DaisyRec}, an open-source Python Toolkit: \url{https://github.com/AmazingDD/daisyRec}.}} Matrix factorization views user-item ratings as a matrix with missing values. By factorizing the matrix with the known values, it recovers the missing values as predictions. Weighted Regularized MF~\cite{hu2008collaborative} assigns different weights to the values in the matrix.
Factorization machines~\cite{rendle2010factorization} consider additional second-order feature interactions of users and items.
Neural MF~\cite{he2017neural} is a combination of generalized MF (GMF) and a multilayer perceptron (MLP). 
Hyper-parameter settings: The number of factors is 200. Regularization weight is 0.0001. We run for 50 epochs with a learning rate of 0.01 with the exception of \textsf{MI} that uses a learning rate of 0.02 for MF and FM. The dropout of NeuMF is set to 0.2.
\paragraph{ConvMF} A CNN-based model proposed by~\citet{kim2016convolutional}\footnote{\url{https://github.com/cartopy/ConvMF}.} that utilizes a convolutional neural network (CNN) for feature encoding of text embeddings. Hyper-parameter settings: The regularization factor is 10 for the user model and 100 for the item model. We used a dropout rate of 0.2.
\paragraph{ANR} Aspect-based Neural Recommender~\cite{chin2018anr}\footnote{\url{https://github.com/almightyGOSU/ANR}.} first proposes aspect-level representations of reviews but its aspects are completely latent without constraints or definitions on the semantics. Hyper-parameter settings: $L_2$ regularization is \num{1e-6}. Learning rate is 0.002. Dropout rate is 0.5. We used 300-dimensional pretrained Google News word embeddings.
\paragraph{DeepCoNN} DeepCoNN~\cite{zheng2017joint}\footnote{Source code of DeepCoNN and NARRE: \url{https://github.com/chenchongthu}.} separately encodes user reviews and item reviews by complex neural networks. Hyper-parameter settings: Learning rate is 0.002 and dropout rate is 0.5. Word embedding is the same as ANR.
\paragraph{NARRE} A model similar to DeepCoNN enhanced by attention mechanism~\cite{chen2018neural}. Attentional weights are assigned to each review to measure its importance.
Hyper-parameter settings: $L_2$ regularization weight is 0.001 Learning rate is 0.002. Dropout rate is 0.5. We used the same word embeddings as described for ANR.
\paragraph{D-Attn\footnote{Source code of D-Attn, MPCN, and DAML: \url{https://github.com/ShomyLiu/Neu-Review-Rec}}} Dual attention-based model~\cite{seo2017interpretable} utilizes CNN as text encoders and builds local- and global-attention (dual attention) for user and item reviews. Hyper-parameter settings: In accordance with the paper, we used 100-dimensional word embedding. The factor number is 200. Dropout rate is 0.5. Learning rate and regularization weight are both 0.001.
\paragraph{MPCN} Multi-Pointer Co-Attention Network~\cite{tay2018multi} selects a useful subset of reviews by pointer networks to build the user profile for the current item. Hyper-parameter settings are the same as D-Attn except that the dropout is 0.2.
\paragraph{DAML} DAML~\cite{liu2019daml} forces encoders of the user and item reviews to interchange information in the fusion layer with local- and mutual- attention so that the encoders can mutually guide the representation generation. Hyper-parameter settings are the same as MPCN.
\paragraph{AHN} Asymmetrical Hierarchical Networks~\cite{dong2020asymmetrical}\footnote{\url{https://github.com/Moonet/AHN}} that guide the user representation generation using item side asymmetric attentive modules so that only relevant targets are significant. Experiments are reproduced following the settings in the paper.

\subsubsection{\rmn{}: Additional Analyses on Hyper-parameter Sensitivity}
\begin{figure}
        \centering
        \begin{subfigure}[b]{0.475\linewidth}
            \centering
            \includegraphics[width=\linewidth]{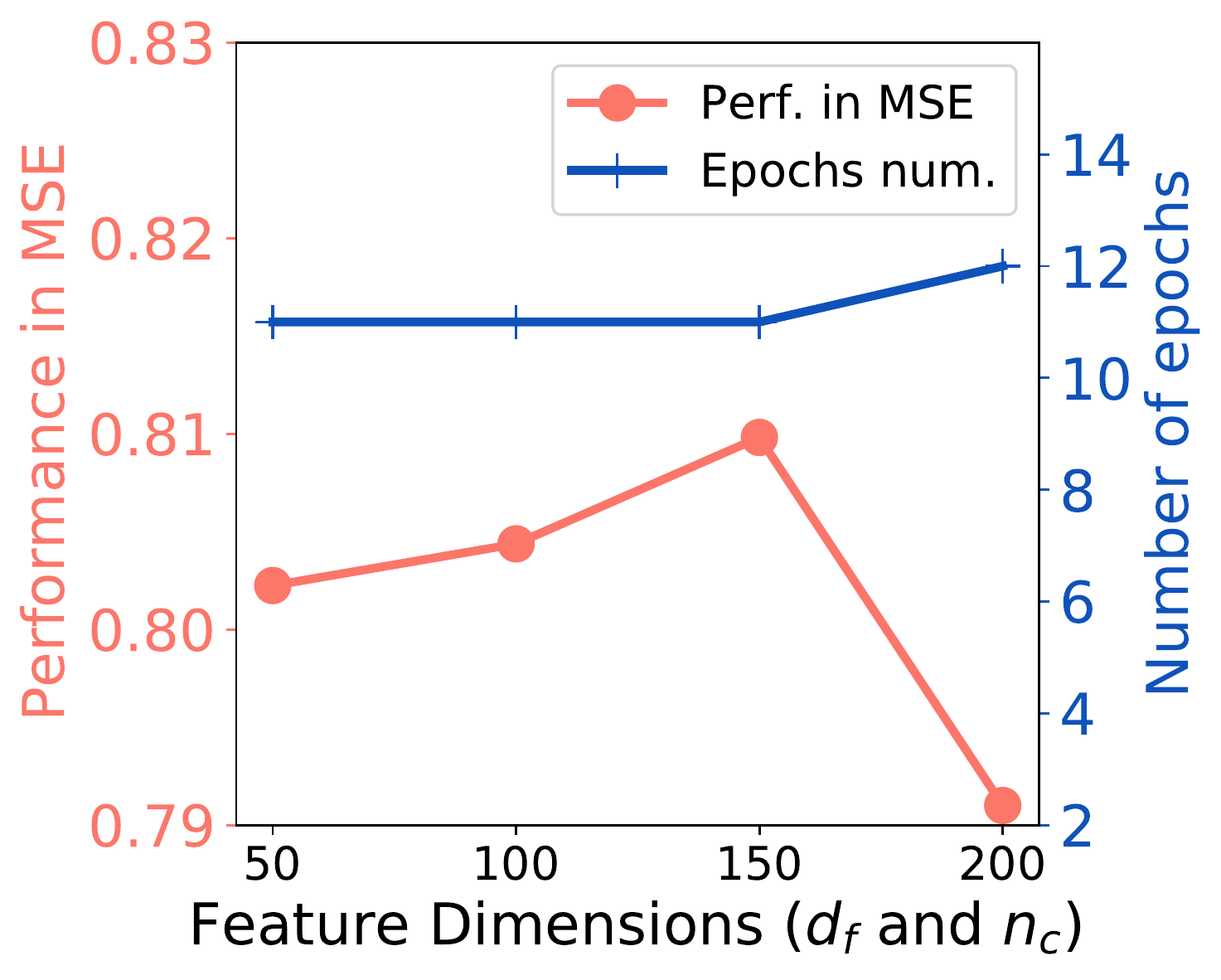}
            \caption{Dims vs. MSE and Ep.}    
            \label{fig:ep_vs_figdim}
        \end{subfigure}%
        ~
        \begin{subfigure}[b]{0.475\linewidth}  
            \centering 
            \includegraphics[width=\linewidth]{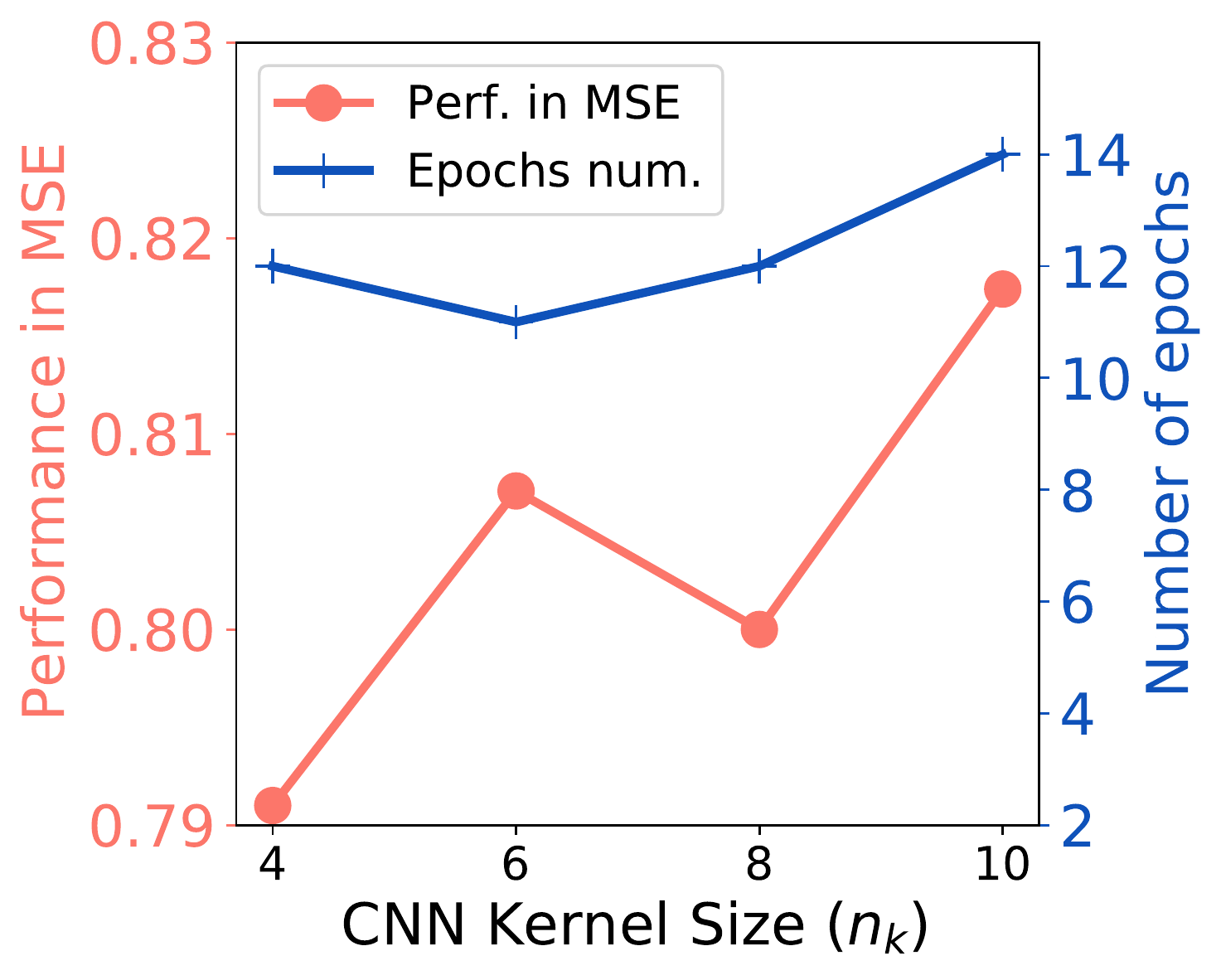}
            \caption{$n_k$ vs. MSE and Ep.}
            \label{fig:ep_vs_kernel}
        \end{subfigure}
        \begin{subfigure}[b]{0.475\linewidth}   
            \centering 
            \includegraphics[width=\linewidth]{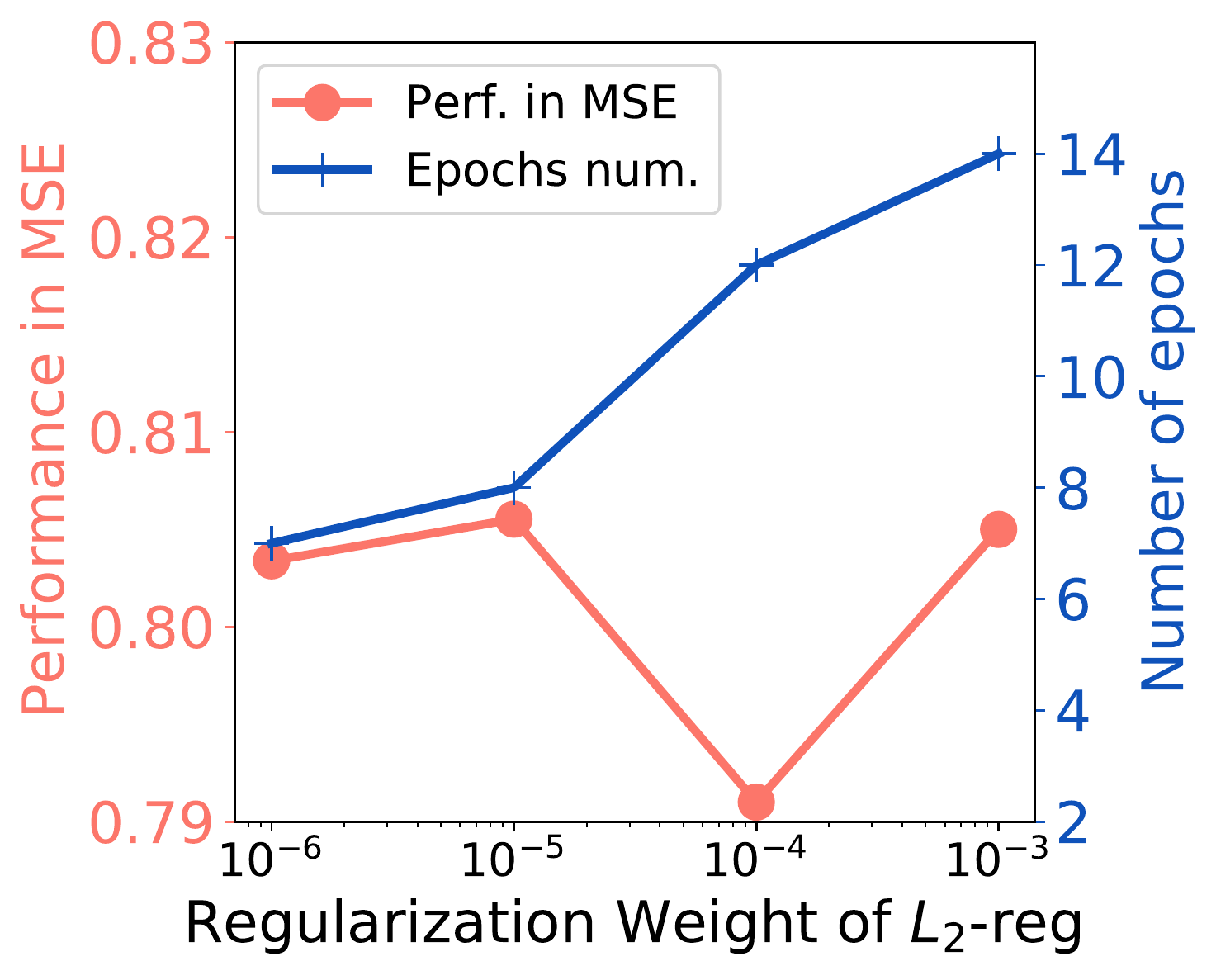}
            \caption{$L_2$-reg vs. MSE and Ep.}    
            \label{fig:ep_vs_regw}
        \end{subfigure}%
        ~
        \begin{subfigure}[b]{0.475\linewidth}   
            \centering 
            \includegraphics[width=\linewidth]{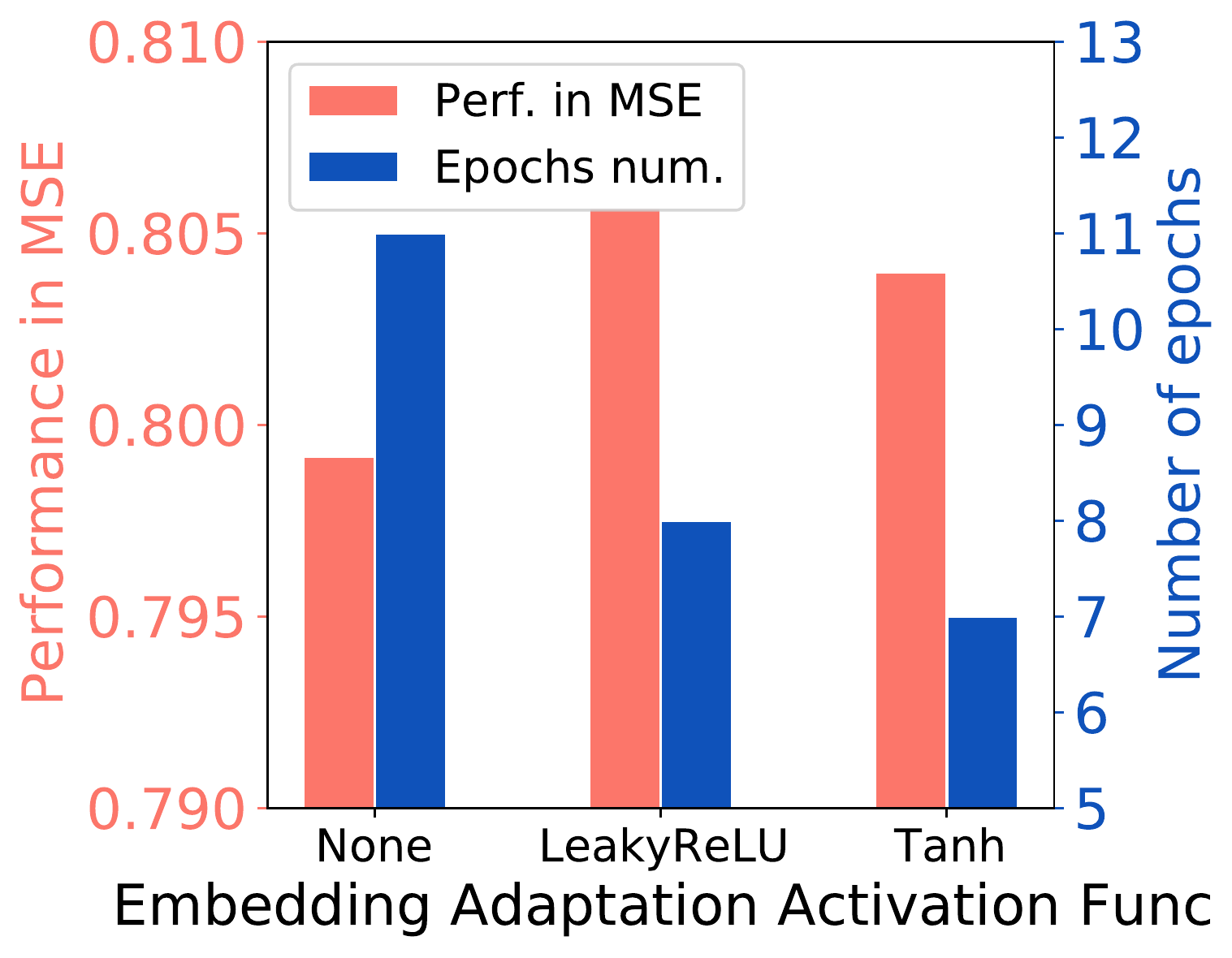}
            \caption{EAF vs. MSE and Ep.}
            \label{fig:ep_vs_activ}
        \end{subfigure}
        \caption{Additional hyper-parameter sensitivity and searching of internal feature dimensions (Dims: $d_f$, $d_a$, and $n_c$), CNN kernel size ($n_k$), regularization weight of $L_2$-reg, and token embedding adaptation function. EAF is short for embedding adaptation function.} 
        \label{fig:add_param_sensitivity}
\end{figure}

Continuing Section~\ref{subsec:rmn}, the searching and sensitivity of the feature dimension ($d_a$, $d_f$, $n_c$), the CNN kernel size $n_k$, and the regularization weight $\lambda$ is exhibited in Figure~\ref{fig:add_param_sensitivity}. 
We always set $d_f=d_a=n_c$ for the consistency of internal feature dimensions. For $(d_f,d_a,n_c)$ in Figure~\ref{fig:ep_vs_figdim}, we choose values from $[50,100,150,200]$ since the output dimension of the BERT encoder is 256. The best performance occurs at 200. The training time spent is stable across different values.
CNN kernel size $n_k$ in Figure~\ref{fig:ep_vs_kernel} varies in $[4,6,8,10]$. We observe that generally larger kernel sizes may in turn hurt the performance as the local features are fused with larger sequential contexts in natural language. The epoch numbers are stable as well.
Figure~\ref{fig:ep_vs_regw} demonstrates how $\lambda$ affects the performance. As $\lambda$ becomes larger, the ``resistance'' against the loss minimization increases so that the training epoch number increases. However, there are no clear trends of performance fluctuation meaning that the sensitivity to $L_2$-reg weight is insignificant.

Finally, we evaluate the effect of adding non-linearity to embedding adaptation function (EAF) mentioned in Section~\ref{subsec:rmn} which transforms $\matH^0$ to $\matH^1$ by $\vech^1_i=\sigma\left(\matW^T_\text{ad}\vech^0_i+\vecb_\text{ad}\right)$. We try LeakyReLU, tanh, and identity functions for $\sigma(\cdot)$ and report the performances in Figure~\ref{fig:ep_vs_activ}.
Without non-linear layers, \rmn{} is able to achieve the best results whereas non-linearity speeds up the training.

\newcommand{\RNum}[1]{\uppercase\expandafter{\romannumeral #1\relax}}
\subsubsection{Case Study \RNum{2} for Interpretation}
\label{subsec:case_study_ii}
Finally, we show another case study from \textsf{AM} dataset using the same attention-property-score visualization schema as Section~\ref{sec:case_study}. In this case, our model is predicting the score user $u_*$ will give to a color and clarity compound for vehicle surface $t_*$. The mentioned aspects of $u_*$ and the properties of $t_*$ are given in Table~\ref{tab:case_study2_reviews} including three overlapping aspects (\textit{quality}, \textit{look}, \textit{cleaning}) and one unique aspect of each side (\textit{size} of $u_*$ and \textit{smell} of $t_*$). A summarization table, Table~\ref{tab:case_study2_summary}, shows the summarized attentions and properties, the inferred impacts, and the corresponding score components of $\langle\vecgamma, \mathcal{F}_\text{ex}([\matG_u;\matG_t])\rangle$. 

In this case study, we can observe the interesting phenomenon also exemplified in Table~\ref{tab:review_examples} by the contrast between R1 and R3 that the aspect \textit{look}, which has been mentioned by $u_*$ and reviewed negatively as a property of $t_*$ (\textit{``strange yellow color''}), only produces an inconsiderable bad effect (-0.002) on the final score prediction. This indicates that the imperfect look (or color) of the item, although also mentioned by $u_*$ in his/her reviews, receives little attention from $u_*$ and thus poses a tiny negative impact on the predicted rating decision of the user.
The other two overlapping aspects show intuitive correlations between their inferred impacts and the scores. The unique aspects, \textit{size} and \textit{smell}, have relatively small influences on the prediction because they are either not attended aspects or not mentioned properties.

It is also notable that some sentences that carry strong emotions may contain few explicit sentiment mentions, e.g., ``\textit{But for an all in one cleaner and wax I think this outperforms most.}'' It backs the design of \rmn{} which carefully takes implicit sentiment signals into consideration, and also calls for an advanced way for aspect-based sentiment modeling beyond term level. Different proportions of such sentences in different datasets may account for the inconsistency of better performances between the two variants of the ablation study.


\newcommand{\revUserQuality}{\small{\textbf{[To $t_1$]} As soon as I poured it into the bucket and started getting ready, I can tell the product was already \senti{better} \asp{quality} than my previous washing liquid.}} 
\newcommand{\revUserCleaning}{\small{\textbf{[To $t_2$]} \dots \supp{I was able to dry my car in record time and not have any water marks left on the paint.} I just slide the towel over any parts with water and \supp{it left no trace of water} and a \senti{clean} \asp{shine} to my car. \textbf{[To $t_3$]} I had completely neglected these areas, except for \asp{minor} \senti{cleaning} and protection. \supp{Once I applied it, the difference was night and day!}}}
\newcommand{\revUserLookAppd}{\small{\textbf{[To $t_4$]} I bought [this item] because I had neglected my paint job for too long. \dots it made my black \asp{paint job} look \senti{dull}.}}
\newcommand{\revUserSize}{\small{\textbf{[To $t_6$]} The \asp{size} was \senti{great} as well, \supp{allowing me to get larger areas in an easier amount of time} so that I could wash my car quicker than I have in the past.}}

\newcommand{\revItemQuality}{\small{\textbf{[From $u_1$]} Adding too little soap will increase the tendency \dots This thick, \senti{high} \asp{quality} soap helps prevent against that. (\gcmark) \textbf{[From $u_2$]} \dots Cons: A bit pricey, but \supp{quality matters}, and this product absolutely has it. \supp{Worth every cent for sure! (\gcmark)}}}
\newcommand{\revItemLookAppd}{\small{\textbf{[From $u_3$]} I was a bit \senti{disappointed}. It is a \senti{strange yellow} \asp{color} and it is thick and I personally did not care for the smell. (\xmark)}}
\newcommand{\revItemCleaning}{\small{\textbf{[From $u_4$]} As far as \asp{cleaning power} it does \senti{fairly good}, \dots The best cleaning of a car is in steps, but \supp{for an all in one cleaner and wax I think this outperforms most.} (\gcmark)}}
\newcommand{\revItemSmell}{\small{\textbf{[From $u_5$]} Just giving some useful feedback about the truth behind the product \dots that it \asp{smells} \senti{good}. \textbf{[From $u_6$]} \supp{I believe this preserves the wax layer longer} \dots This is much thicker than the [some brand] soap, and has a very \asp{pleasant} \senti{smell} to it. (\gcmark)}}

\begin{table}[!h]
    \centering
    \resizebox{0.9\linewidth}{!}{
    \begin{tabular}{cQ}
    \toprule
    \multicolumn{2}{c}{\textit{From reviews given by user} $u_*$.} \\
    \midrule
    \rowcolor{yellow!10} quality & \revUserQuality \\
    \midrule
    \rowcolor{yellow!10} look & \revUserLookAppd \\
    \midrule
    \rowcolor{yellow!10} cleaning    & \revUserCleaning \\
    \midrule
    \rowcolor{green!10} size & \revUserSize \\
    \midrule
    \midrule
    \multicolumn{2}{c}{\textit{From reviews received by item} $t_*$.} \\
    \midrule
    \rowcolor{yellow!10} quality & \revItemQuality\\
    \midrule
    \rowcolor{yellow!10} look & \revItemLookAppd \\
    \midrule
    \rowcolor{yellow!10} cleaning  & \revItemCleaning \\
    \midrule
    \rowcolor{cyan!10} smell & \revItemSmell \\
    \bottomrule
    \end{tabular}}
    \caption{Examples of reviews from $u_*$ and to $t_*$ with \asp{Aspect}-\senti{Sentiment} pair mentions as well as \supp{other sentiment evidences} on five example aspects.}
    \label{tab:case_study2_reviews}
\end{table}

\begin{table}[!h]
    \centering
     \resizebox{\linewidth}{!}{
     \setlength\tabcolsep{2pt}
    \begin{tabular}{cccccc}
        \topline
        \rowcolor[gray]{0.92}
         Aspects & size & quality & look & cleaning & smell \\
        \midtopline
        Attn. of $u_*$ & \ccg \bcmark & \ccy \bcmark & \ccy \textbf{--} & \ccy \bcmark & n/a \\
        Prop. of $t_*$ &  n/a  & \ccy \gcmark & \ccy \xmark & \ccy \gcmark & \ccc \gcmark  \\
        Inferred Impact & \unc & \posImp & \negImp & \posImp & \unc\\
        \midrule
        $\vecgamma_i\mathcal{F}_{\text{ex}} (\cdot)_i$ ($\times 10^{-2}$) & \ccgray 0.5 & \ccg 2.9 & \ccr -0.2 & \ccg 1.4 & \ccgray 0.3\\ 
        \bottomrule
    \end{tabular}}
    \caption{Attentions and properties summaries, inferred impacts, and the learned aspect-level contributions on the score prediction.}
    \label{tab:case_study2_summary}
\end{table}


\end{document}